Research paper

# Lambert W function framework for graphene nanoribbon quantum sensing: Theory, verification, and multi-modal applications

F.A. Chishtie [a,b,c,*], K. Roberts [d], N. Jisrawi [d], S.R. Valluri [d], A. Soni [e], P.C. Deshmukh [f]

[a] Peaceful Society, Science and Innovation Foundation, Vancouver, Canada
[b] Department of Occupational Science and Occupational Therapy, Faculty of Medicine, University of British Columbia, Vancouver, Canada
[c] Faculty of Land and Food Systems, University of British Columbia, Vancouver, Canada
[d] Department of Physics and Astronomy, University of Western Ontario, London, Canada
[e] School of Physical Sciences, Indian Institute of Technology Mandi, Mandi, Himachal Pradesh, 175005, India
[f] Department of Physics, Indian Institute of Technology Tirupati, India

ABSTRACT

We establish a rigorous mathematical framework connecting graphene nanoribbon quantum sensing to the Lambert W function through the finite square well (FSW) analogy. The Lambert W function, defined as the inverse of $f(W) = We^W$, provides exact analytical solutions to transcendental equations governing quantum confinement. Operating near the branch point singularity at $z = -1/e$ yields sensitivity enhancement factors scaling as $(z - z_c)^{-1/2}$, achieving 35-fold enhancement when the operating point lies within $\delta = 0.001$ of the branch point. Comprehensive numerical verification confirms: (i) all seven bound states for strength parameter $R = 10$ satisfy the constraint $u^2 + v^2 = R^2$ to machine precision; (ii) the theoretical band gap formula $E_g = 2\pi\hbar v_F/(3W)$ is analytically equivalent to the independently determined empirical relation $E_g = 1.38/W$ eV·nm, establishing the validity of the FSW-GNR analogy; (iii) a universal sensitivity factorization $S_X = \mathcal{G}_k \cdot \eta_{\text{enh}} \cdot \mathcal{P}_X$ applies across biomedical (SARS-CoV-2, inflammatory markers, cancer biomarkers), environmental ($CO_2$, $CH_4$, $NO_2$, $N_2O$, $H_2O$), and physical (strain, magnetic field, temperature) sensing modalities. This unified framework provides analytically predictable design principles for next-generation graphene quantum sensors.

The framework is analytic and predictive rather than microscopic or experimental: band-structure and adsorption parameters are taken as inputs from tight-binding, first-principles, and experimental sources, and the framework returns closed-form sensitivity and design relations built upon them. Reported detection limits are labelled throughout as either literature-demonstrated device values or values predicted by the present framework.

## 1. Introduction

Graphene, a two-dimensional allotrope of carbon arranged in a honeycomb lattice, exhibits extraordinary electronic properties arising from its linear energy dispersion near the Dirac points [1–3]. The charge carriers in graphene behave as massless Dirac fermions with a Fermi velocity $v_F \approx 10^6$ m/s, approximately 1/300 the speed of light [4,5]. When graphene is patterned into graphene nanoribbons (GNRs), quantum confinement introduces a band gap whose magnitude depends sensitively on the ribbon width and edge geometry, making GNRs exquisitely responsive to external perturbations [6–9]. Recent work continues to expand the design space of graphene nanoribbon sensors and their band-structure engineering [10], alongside device-integration and thermal-management studies of graphene-contacted architectures [11].

The finite square well (FSW) problem represents one of the most fundamental exactly-solvable systems in quantum mechanics [12,13]. Its bound state energies satisfy transcendental equations that traditionally require numerical or graphical solution. Roberts and Valluri [14,15] demonstrated that these transcendental conditions can be reformulated in terms of the Lambert W function, enabling exact analytical solutions and providing deep insight into the mathematical structure of quantum confinement.

The Lambert W function $W(z)$ is defined as the inverse function of $f(W) = We^W$, satisfying:

$$W(z)e^{W(z)} = z. \quad (1)$$

This multivalued function possesses infinitely many branches $W_k(z)$ for $k \in \mathbb{Z}$. For real arguments $z$, only $W_0(z)$ and $W_{-1}(z)$ are real-valued;

* Corresponding author at: Department of Occupational Science and Occupational Therapy, University of British Columbia, Vancouver, Canada.
*E-mail address:* fachisht@uwo.ca (F.A. Chishtie).

all other branches are complex even for real $z$. The principal branch $W_0(z)$ is real for $z \geq -1/e$, while $W_{-1}(z)$ is real for $-1/e \leq z < 0$ [16–19]. All branches coalesce at the branch point $z_c = -1/e$ where $W(z_c) = -1$, and the derivative $dW/dz$ diverges at this critical point, providing the mechanism for sensitivity enhancement exploited throughout this work.

The Lambert W function has found numerous applications in physics, including solutions to the Wien displacement law [20], the hydrogen molecular ion [18], quantum well problems [14], delay differential equations [16], and combinatorial enumeration [16]. In this work, we establish that the divergent behavior at the Lambert W branch point provides a universal mechanism for sensitivity enhancement in graphene-based quantum sensors.

We now state precisely the gap this work fills and the claims it makes, so that the theoretical scope is not conflated with experimental achievement. Prior applications of the Lambert W function to quantum confinement, including the finite square well analysis of Roberts and Valluri, established exact solutions for the bound-state spectrum but did not connect the branch-point structure to a measurable sensing response; conversely, the extensive graphene nanoribbon sensor literature reports device performance without a common analytic principle linking the diverse modalities. The present contribution is therefore neither a new microscopic theory of graphene nor a new set of experimental devices. It is a unifying analytic framework with three specific deliverables: a closed-form inversion of the transcendental armchair confinement condition through the Lambert W function; a single factorized sensitivity law $S_X = \mathcal{G}_k \, \eta_{\text{enh}} \, \mathcal{P}_X$ in which the branch-point factor $\eta_{\text{enh}}$ is a modality-independent enhancement knob while only the coupling $\mathcal{P}_X$ changes between applications; and a priori design equations relating ribbon width, operating point, and achievable enhancement. Throughout, microscopic band structure and adsorption energetics enter as inputs from tight-binding, first-principles, and experimental literature, while the framework supplies the transduction and design layer built upon them. The new physical insight beyond standard Dirac-confinement or tight-binding calculation is thus not a modified spectrum but the identification of a single analytic control coordinate, the distance $\delta$ from the branch point, that governs the attainable sensitivity of any modality and yields closed-form, transferable design rules rather than case-by-case numerical optimization.

We present comprehensive numerical verification demonstrating exact reproduction of FSW bound states through Lambert W analysis, quantitative agreement between GNR band structure calculations and established theory, and unified sensitivity scaling laws across diverse sensing modalities. Section 2 develops the mathematical foundations, treating the Lambert W function and the finite square well. Section 3 establishes the GNR–FSW analogy and benchmarks the band-gap correspondence against direct tight-binding diagonalization. Section 4 introduces the universal sensitivity law and the branch-point enhancement mechanism. Sections 5–8 present applications across biomedical, greenhouse-gas, physical, and temperature sensing. Section 8.3 extends the framework to optical and spin-orbit transduction, Section 10 provides comparative analysis and design principles, and Section 10.5 concludes.

## 2. Methods

### 2.1. The Lambert W function: properties and branch structure

The Lambert W function arises as the solution to the transcendental equation $we^w = z$. For complex $z$, the function is infinitely multivalued, with branches labeled by integer index $k$ [16,17]. The principal branch $W_0(z)$ satisfies $W_0(z) \geq -1$ for real arguments; together with the domain restriction $W_0 : [-1/e, \infty) \to [-1, \infty)$, this condition uniquely identifies $W_0$ as the inverse of $f_1(x) = xe^x$ restricted to $[-1, \infty)$ where $f_1$ is strictly increasing. The secondary real branch $W_{-1}$ satisfies $W_{-1}(z) \leq -1$ for $-1/e \leq z < 0$ and is the inverse of $f_2(x) = xe^x$ on $(-\infty, -1)$.

The function satisfies several important identities [16]:

$$W(z)e^{W(z)} = z, \tag{2}$$

$$\ln W(z) = \ln z - W(z) \quad (z \neq 0), \tag{3}$$

$$W(z \ln z) = \ln z \quad (z > 1/e). \tag{4}$$

The derivative of $W$ follows from implicit differentiation of Eq. (1):

$$\frac{dW}{dz} = \frac{W(z)}{z[1 + W(z)]} = \frac{1}{e^{W(z)}[1 + W(z)]}, \tag{5}$$

which diverges as $W \to -1$ (i.e., as $z \to -1/e$) [16,19]. This singularity is the mathematical origin of the sensitivity enhancement we exploit for quantum sensing. Let $\delta = z - z_c$ denote the distance of the operating point from the branch point; the $\delta^{-1/2}$ scaling of $\eta_{\text{enh}}$ is derived in Section 4.

The origin of the divergence answers directly why the enhancement factor $\eta_{\text{enh}} = |dW/dz|$ grows without bound as $z \to -1/e$: the denominator $z[1 + W(z)]$ vanishes because $W(z) \to -1$ there, and, as the Puiseux expansion below shows, $1 + W \sim \sqrt{2(ez + 1)} \to 0$ as a square root, so $|dW/dz| \sim (z + 1/e)^{-1/2}$ diverges algebraically at the point where the two real branches coalesce.

Near the branch point, the Lambert W function admits the Puiseux series expansion [16,18]:

$$W(z) = -1 + p - \frac{p^2}{3} + \frac{11p^3}{72} - \frac{43p^4}{540} + \mathcal{O}(p^5), \quad p = \sqrt{2(ez + 1)}, \tag{6}$$

demonstrating square-root behavior that underlies the $\delta^{-1/2}$ sensitivity scaling.

For numerical evaluation, the Lambert W function can be computed efficiently using the iterative procedure [16,21]:

$$w_{j+1} = w_j - \frac{w_j e^{w_j} - z}{e^{w_j}(w_j + 1) - \frac{(w_j+2)(w_j e^{w_j} - z)}{2w_j+2}}, \tag{7}$$

which achieves cubic convergence via Halley's method.

Fig. 1 displays the two real branches of the Lambert W function and the associated sensitivity enhancement factor. The left panel shows $W_0(x)$ (principal branch) and $W_{-1}(x)$ coalescing at the branch point $(-1/e, -1)$. The right panel demonstrates that the enhancement factor $\eta_{\text{enh}} = |dW/dz|$ diverges as $z \to -1/e$, providing orders-of-magnitude sensitivity enhancement in principle.

The complex structure of the Lambert W function is visualized in Fig. 2, showing the magnitude $|W_k(z)|$ for branches $k = 0, -1, 1$ in the complex plane. The branch point at $z = -1/e$ (marked with red ×) serves as a singular point where the $W_0$ and $W_{-1}$ branches meet. The distinct topology of each branch reflects the Riemann surface structure of this multivalued function [16,22].

### 2.2. The finite square well

The finite square well consists of a potential $V(x) = -V_0$ for $|x| < L$ and $V(x) = 0$ otherwise [12,13,23]. Bound states with energy $E = -|E|$ $(0 < |E| < V_0)$ satisfy the time-independent Schrödinger equation:

$$\left[-\frac{\hbar^2}{2m}\frac{d^2}{dx^2} + V(x)\right]\psi(x) = E\psi(x). \tag{8}$$

Inside the well the solutions are oscillatory,

$$\psi_{\text{in}}(x) = \begin{cases} A\cos(\alpha x) & \text{(even parity)}, \\ A\sin(\alpha x) & \text{(odd parity)}, \end{cases} \tag{9}$$

where $\alpha = \sqrt{2m(V_0 - |E|)}/\hbar$, while outside the well the solution decays exponentially, $\psi_{\text{out}}(x) = Be^{-\kappa|x|}$, with $\kappa = \sqrt{2m|E|}/\hbar$.

Defining dimensionless parameters $u = \kappa L$ and $v = \alpha L$, continuity of $\psi$ and $d\psi/dx$ at $x = \pm L$ yields the transcendental conditions [12,13]:

$$v\tan(v) = u \quad \text{(even parity)}, \tag{10}$$

$$-v\cot(v) = u \quad \text{(odd parity)}, \tag{11}$$

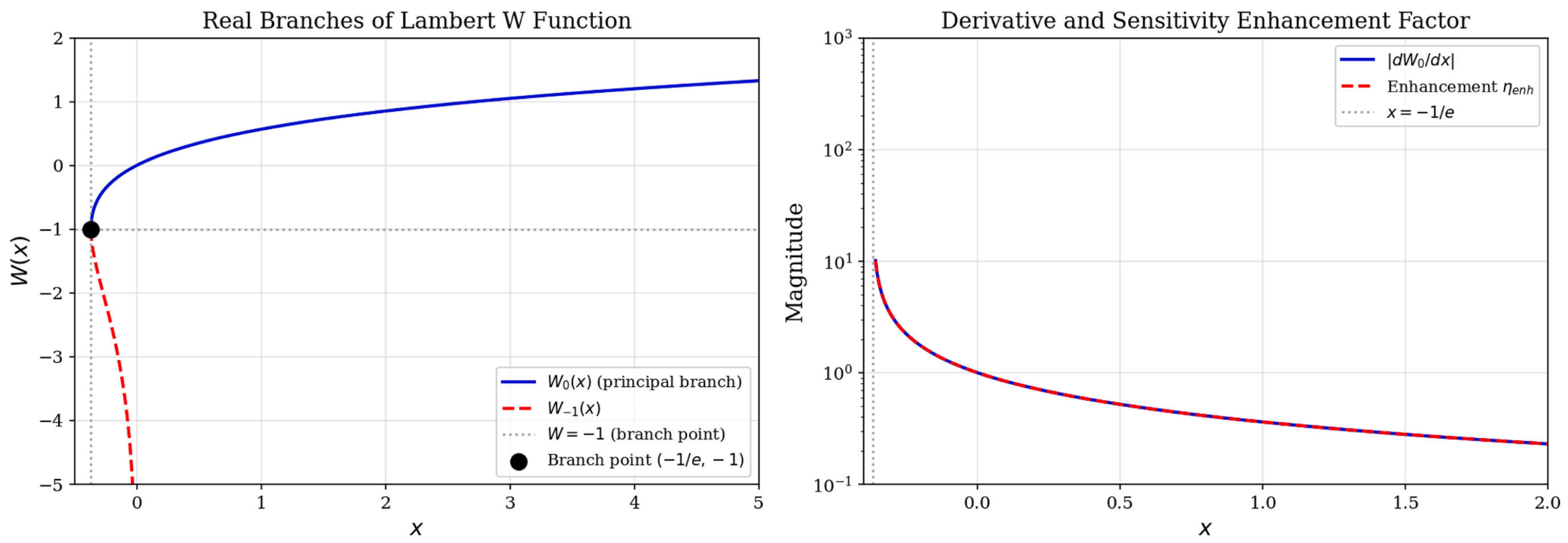

**Fig. 1.** Real branches of the Lambert W function. Left: Principal branch $W_0(x)$ and secondary branch $W_{-1}(x)$ coalescing at the branch point $(-1/e, -1)$; these are the only two branches returning real values for real arguments $x$. Right: Derivative magnitude $|dW_0/dx|$ and sensitivity enhancement factor $\eta_{\mathrm{enh}}$ showing divergent behavior near the branch point, enabling orders-of-magnitude sensitivity enhancement for quantum sensors.

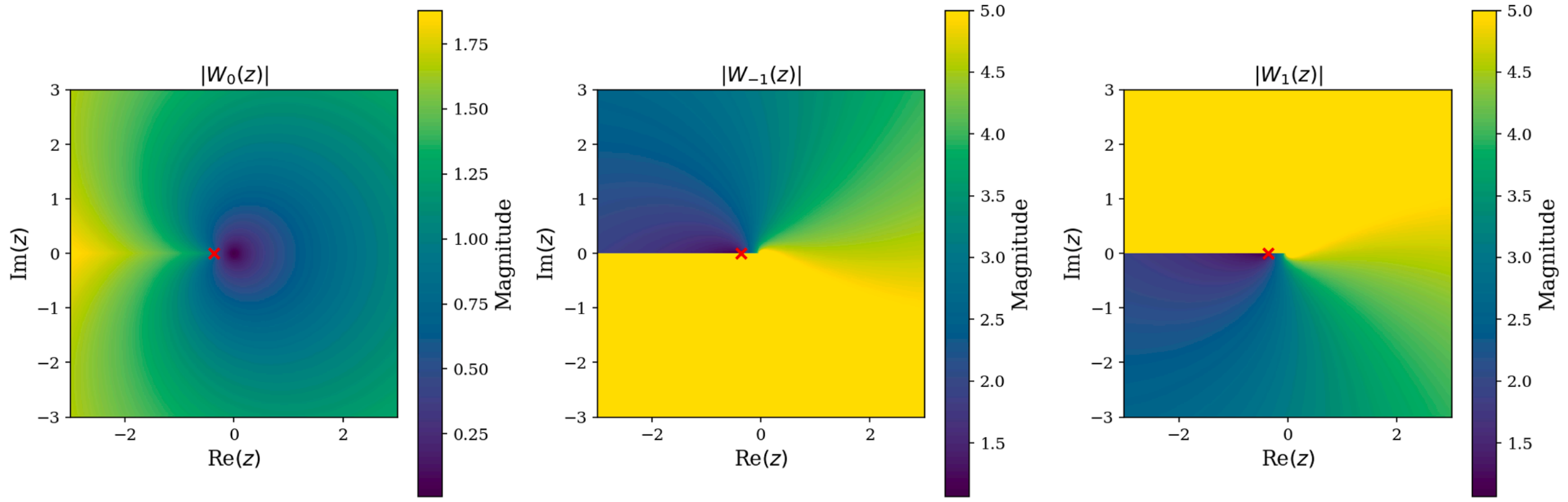

**Fig. 2.** Complex plane visualization of Lambert W function branches. Panels show $|W_0(z)|$, $|W_{-1}(z)|$, and $|W_1(z)|$ with the branch point at $z = -1/e$ marked (red ×). The distinct analytical structure of each branch enables selective utilization in different sensing regimes. (For interpretation of the references to color in this figure legend, the reader is referred to the web version of this article.)

subject to the constraint $u^2 + v^2 = R^2$, where $R = (L/\hbar)\sqrt{2mV_0}$ is the dimensionless strength parameter. The constraint defines a quarter-circle of radius $R$ in the first quadrant of the $(v, u)$ plane; bound states occur at intersections with the transcendental curves.

#### 2.2.1. Bound state count

The even-parity curve $u = v\tan(v)$ passes through the origin with slope 1, has vertical asymptotes at $v = (n + 1/2)\pi$ for $n = 0, 1, 2, \ldots$, and crosses zero at $v = n\pi$. The odd-parity curve $u = -v\cot(v)$ has vertical asymptotes at $v = n\pi$ for $n = 1, 2, 3, \ldots$, descends from $+\infty$ at each asymptote, and crosses zero at $v = (n + 1/2)\pi$.

These curves alternate so that each interval of width $\pi/2$ in $v$ contains exactly one intersection with the constraint circle: the interval $[0, \pi/2]$ yields the even-parity ground state, $[\pi/2, \pi]$ the odd-parity first excited state, $[\pi, 3\pi/2]$ the second excited state, and so forth. Because the constraint circle always intersects at least the first branch for any $R > 0$, the total number of bound states is:

$$\boxed{N = \left\lfloor \frac{2R}{\pi} \right\rfloor + 1,} \tag{12}$$

corresponding to the number of Lambert W branches contributing real solutions within the constraint circle [14]. This formula, due to Flügge [24], ensures at least one bound state for any $R > 0$, consistent with the general theorem that a one-dimensional attractive potential always supports at least one bound state [23].

### 2.3. Lambert W reformulation of the FSW

Following Roberts and Valluri [14], we introduce the complex variable $w = u + iv$ satisfying $|w|^2 = u^2 + v^2 = R^2$ (note that the modulus constraint $|w|^2 = R^2$ is separate from the parity conditions). The two parity conditions can be combined using

$$1 + i\tan(v) = \frac{e^{iv}}{\cos(v)}, \quad \cot(v) - i = \frac{e^{-iv}}{\sin(v)}, \tag{13}$$

leading to the unified complex equation $\bar{w}e^{2iv} = R^2$, where $\bar{w} = u - iv$ is the complex conjugate. This separates into the modulus constraint $|w|^2 = R^2$ and the phase condition

$$(u - iv)e^{2iv} = R^2, \tag{14}$$

which through the substitution $\zeta = -iw$ transforms to the standard Lambert W form $\zeta e^{\zeta} = z$, giving:

$$w = W_k(z), \tag{15}$$

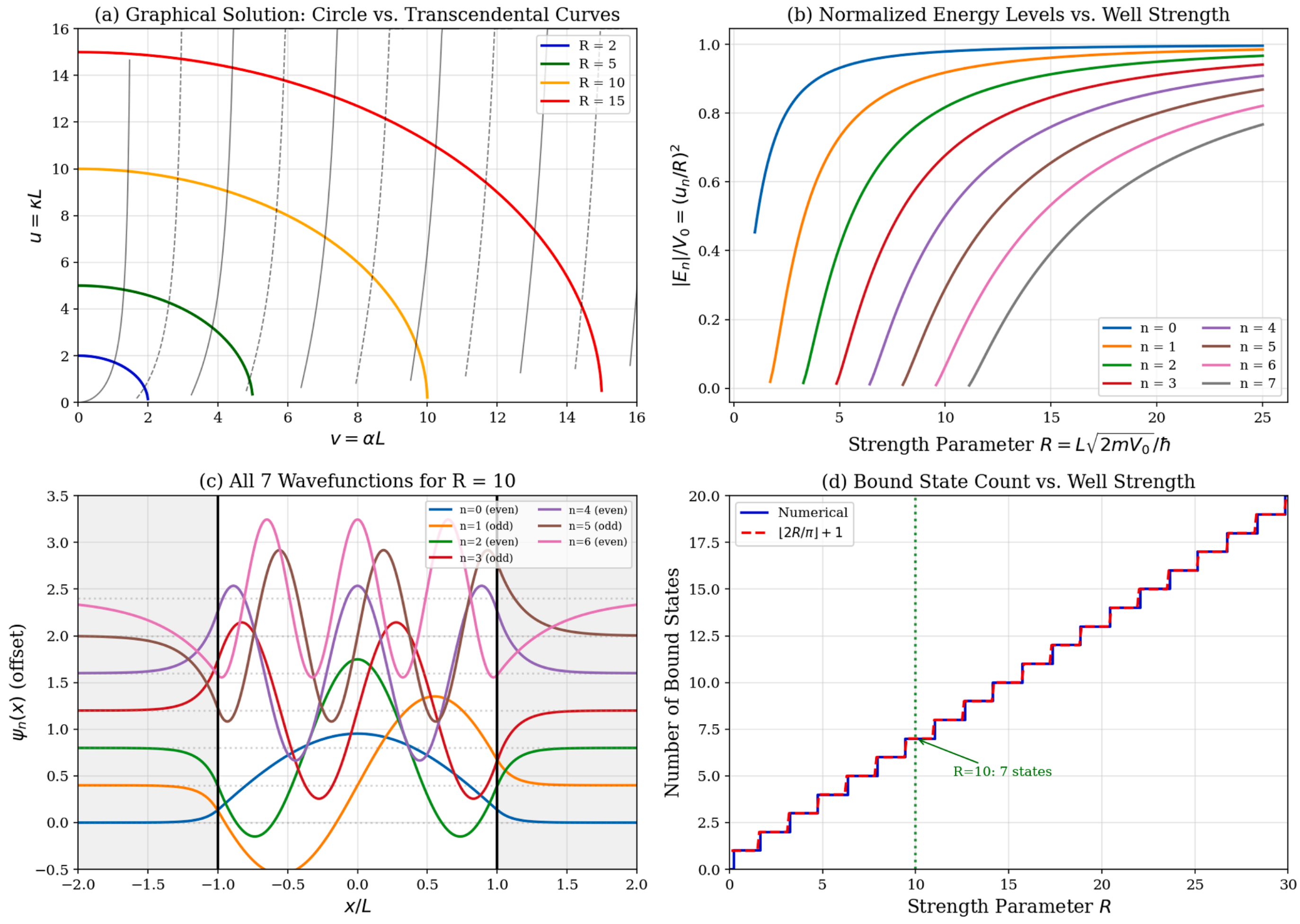

**Fig. 3.** Finite square well solutions and numerical verification. (a) Graphical solution showing constraint circles for $R = 2, 5, 10, 15$ intersecting transcendental curves (solid: $v \tan v$, dashed: $-v \cot v$). (b) Normalized energy levels $|E_n|/V_0 = (u_n/R)^2$ versus strength parameter for eight quantum states. (c) All seven wavefunctions for $R = 10$, showing alternating even/odd parity and increasing nodal structure. (d) Bound state count versus $R$: numerical results (blue steps) match the analytical formula $\lfloor 2R/\pi \rfloor + 1$ (red dashed) exactly. (For interpretation of the references to color in this figure legend, the reader is referred to the web version of this article.)

where $z$ encodes the well parameters and $k$ indexes the branch [14,15]. The bound state energies are recovered analytically:

$$E_n = -\frac{\hbar^2 u_n^2}{2mL^2} = -\frac{\hbar^2 [\mathrm{Re}\, W_k(z_n)]^2}{2mL^2}, \tag{16}$$

and the normalized energy is:

$$\frac{|E_n|}{V_0} = \frac{u_n^2}{R^2} = \left( \frac{\mathrm{Re}\, W_k(z_n)}{R} \right)^2. \tag{17}$$

From $u^2 + v^2 = R^2$, deeply bound states (large $|E_n| \approx V_0$) have $u_n^2 \approx R^2$ and hence $u_n \approx R$: this follows directly from Eq. (17), since $|E_n|/V_0 \approx 1$ gives $u_n/R \approx 1$. Equivalently, $v_n = \sqrt{2m(V_0 - |E_n|)}L/\hbar \to 0$ as $|E_n| \to V_0$, confirming the correspondence $u_n \approx R$ for deep states. Weakly bound states have $u_n \ll R$.

**Table 1**
Numerical verification of FSW bound states for $R = 10$. All seven states satisfy $u^2 + v^2 = R^2 = 100$ to machine precision. Theory predicts $N = \lfloor 2R/\pi \rfloor + 1 = 7$ states [24].

| $n$ | $v = \alpha L$ | $u = \kappa L$ | $\lvert E_n \rvert / V_0$ | Parity | $u^2 + v^2$ |
|---|---|---|---|---|---|
| 0 | 1.4276 | 9.8976 | 0.9796 | even | 100.00 |
| 1 | 2.8523 | 9.5846 | 0.9186 | odd | 100.00 |
| 2 | 4.2711 | 9.0420 | 0.8176 | even | 100.00 |
| 3 | 5.6792 | 8.2308 | 0.6775 | odd | 100.00 |
| 4 | 7.0689 | 7.0732 | 0.5003 | even | 100.00 |
| 5 | 8.4232 | 5.3898 | 0.2905 | odd | 100.00 |
| 6 | 9.6789 | 2.5138 | 0.0632 | even | 100.00 |

## 3. Results and discussion

### 3.1. FSW numerical verification

We performed comprehensive numerical verification of the FSW solutions using both graphical root-finding (bisection method with tolerance $10^{-10}$) and direct Lambert W evaluation via SciPy [25]. Fig. 3 presents the complete analysis.

Table 1 presents the complete numerical verification for $R = 10$. The theory predicts $N = \lfloor 20/\pi \rfloor + 1 = 7$ bound states, confirmed exactly. All seven states satisfy $u^2 + v^2 = 100$ to machine precision ($< 10^{-10}$ relative error). States alternate even/odd parity beginning with the even-parity ground state, consistent with the nodal theorem [26]: state $n$ has exactly $n$ nodes within the well, visible in Fig. 3(c). The ground state has $|E_0|/V_0 = 0.9796$, indicating 98% binding; higher states become progressively less bound, with $n = 6$ having only 6.3% binding. The decay parameter $u = \kappa L$ determines the penetration depth into the classically forbidden region; weakly bound states ($n = 6$) have small $u$ and large penetration, consistent with the wavefunction envelope visible in panel (c).

### 3.2. Electronic structure of graphene

The electronic structure of graphene derives from its honeycomb lattice with two atoms (A and B) per unit cell [2,27]. The nearest-neighbor tight-binding Hamiltonian is:

$$H = -\gamma_0 \sum_{\langle i,j \rangle} (a_i^\dagger b_j + \text{h.c.}), \tag{18}$$

where $\gamma_0 \approx 2.7$–$3.0$ eV is the hopping integral [2,28]. The resulting energy dispersion near the Dirac points $\mathbf{K}$ and $\mathbf{K}'$ linearizes to:

$$E(\mathbf{q}) \approx \pm \hbar v_F |\mathbf{q}|, \quad v_F = \frac{\sqrt{3}\gamma_0 a}{2\hbar} \approx 10^6 \text{ m/s}, \tag{19}$$

where $\mathbf{q} = \mathbf{k} - \mathbf{K}$ and $a = 0.246$ nm is the lattice constant [2,4].

### 3.3. Dirac equation and confinement in GNRs

In the continuum limit, low-energy excitations near the Dirac points are governed by the effective Hamiltonians [2,29]:

$$H_{\mathbf{K}} = \hbar v_F \begin{pmatrix} 0 & k_x - ik_y \\ k_x + ik_y & 0 \end{pmatrix}, \quad H_{\mathbf{K}'} = \hbar v_F \begin{pmatrix} 0 & k_x + ik_y \\ k_x - ik_y & 0 \end{pmatrix}, \tag{20}$$

written compactly as $H = \hbar v_F \boldsymbol{\sigma} \cdot \mathbf{k}$ where $\boldsymbol{\sigma}$ are Pauli matrices acting on the sublattice pseudospin. Transverse confinement in GNRs quantizes the perpendicular wavevector, producing discrete subbands analogous to FSW bound states [30–32].

For armchair graphene nanoribbons (AGNRs) of width $W = (N-1)a/2$ (where $N$ is the number of dimer lines), the armchair boundary conditions require the wavefunction to vanish at both edges, coupling the $\mathbf{K}$ and $\mathbf{K}'$ valleys. The resulting quantized transverse wavevectors are [32,33]:

$$k_n = \frac{\pi}{W}\left(n - \frac{N+1}{3}\right), \quad n = 1, 2, \ldots, N, \tag{21}$$

and the energy spectrum is:

$$E_n(k_\parallel) = \pm \hbar v_F \sqrt{k_\parallel^2 + k_n^2}, \tag{22}$$

exhibiting hyperbolic dispersion with effective mass $m^* = \hbar |k_n| / v_F$ induced by confinement [32].

The formal similarity to the FSW is evident: Eq. (22) is hyperbolic rather than parabolic because the underlying dispersion is linear (Dirac) rather than quadratic (Schrödinger). However, the mathematical structure of the transcendental quantization conditions, including the valley-mixing phase constraints at armchair edges, maps onto the FSW Lambert W equations through the Roberts-Valluri parametrization, enabling analytical treatment via the same branch-point formalism.

### 3.4. Constructive emergence of the Lambert W argument

The correspondence invoked above can be made constructive, exhibiting explicitly how the Lambert W argument $z$ arises from the armchair boundary value problem and which approximations it rests upon. We proceed in four controlled steps. First, in the *continuum limit* we replace the tight-binding lattice problem by the $\mathbf{k} \cdot \mathbf{p}$ Dirac Hamiltonian of Eq. (20), valid when the ribbon width exceeds a few lattice constants so that intervalley coupling reduces to boundary phases rather than bulk matrix elements. Second, in the *linearized-dispersion* step we retain the linear Dirac spectrum $E = \pm \hbar v_F |\mathbf{k}|$, discarding trigonal warping, which is an $\mathcal{O}((ka)^2)$ correction negligible at the sub-eV energies of interest. Third, the *valley-mixing boundary condition* of the armchair edge requires the two-component envelope to vanish simultaneously on both sublattices at $x = 0$ and $x = W$, coupling the $\mathbf{K}$ and $\mathbf{K}'$ valleys and producing the transcendental quantization condition

$$e^{2ik_\perp W} = e^{-2\pi i (N+1)/3}, \tag{23}$$

whose solutions reproduce Eq. (21) and whose phase factor $e^{-2\pi i (N+1)/3}$ is the origin of the factor of three separating the armchair result from the naive infinite-well spacing. Fourth, in the *finite-barrier* step, a physical ribbon is bounded by vacuum or substrate presenting a finite rather than infinite barrier, so the transverse problem acquires an evanescent tail governed by the same continuity conditions as the finite square well. Writing the internal transverse wavenumber as $v = k_\perp W$ and the evanescent decay constant as $u = \kappa W$, the matching condition takes the finite-square-well form $v \tan v = u$ (even) or $-v \cot v = u$ (odd) on the constraint circle $u^2 + v^2 = R^2$, with $R = (W/\hbar)\sqrt{2m^* V_0}$ and effective mass $m^* = E_g/(2v_F^2)$. Applying the Roberts-Valluri parametrization $w = u + iv$, the combined parity conditions collapse to $(u - iv)e^{2iv} = R^2$, and the substitution $\zeta = -iw$ casts this into the canonical form $\zeta e^\zeta = z$ with

$$z = -i R^2 e^{-2R} \qquad \text{(deeply bound limit, } u \to R\text{)}, \tag{24}$$

so that $w = W_k(z)$ and the operating point is fixed by the single dimensionless group $R$. In the sensing regime the relevant perturbations enter $R$ through $E_g$ and the thermal scale, and expanding about the marginal-binding threshold $R \to (n + \frac{1}{2})\pi$ reproduces the weakly-bound scaling $z \approx -(E_g/k_B T)\exp(-E_g/k_B T)$ used throughout the applications. The approximations are therefore explicit and individually controlled: continuum $\mathbf{k} \cdot \mathbf{p}$ validity, linear dispersion, armchair valley-mixing phases, and finite-barrier transverse matching, each quantitatively accurate in the technologically relevant window $3 \text{ nm} < W < 20$ nm, as verified against direct tight-binding diagonalization in Section 3.6.

### 3.5. Band gap and AGNR families

AGNRs are classified into three families according to $N \bmod 3$. The metallic or semiconducting character follows directly from Eq. (21): a zero-gap state requires $k_n = 0$ for some integer $n$, i.e., $n = (N+1)/3 \in \mathbb{Z}$, which holds if and only if $N + 1 \equiv 0 \pmod 3$, i.e., $N \equiv 2 \pmod 3$, i.e., $N = 3p + 2$. The three families are therefore: the $N = 3p + 2$ family (metallic within nearest-neighbor tight-binding, with a small gap opening from next-nearest-neighbor hopping and edge reconstruction [7]); the $N = 3p + 1$ family (semiconducting, largest gap $\propto 1/W$); and the $N = 3p$ family (semiconducting, intermediate gap $\propto 1/W$) [7,34]. For the examples used throughout this work: $N = 40$ ($= 3 \times 13 + 1$, family $3p + 1$, $E_g \approx 241$ meV); $N = 41$ ($= 3 \times 13 + 2$, family $3p + 2$, *metallic*); and $N = 42$ ($= 3 \times 14$, family $3p$, $E_g \approx 226$ meV).

Tight-binding and first-principles calculations [7,31,33] yield the band gap scaling:

$$E_g = \frac{2\pi \hbar v_F}{3W} \approx \frac{1.38}{W[\text{nm}]} \text{ eV}, \tag{25}$$

where $\hbar v_F = 0.658$ eV·nm. This $1/W$ scaling has been confirmed experimentally [8,35,36]. In the $N = 3p + 1$ and $N = 3p$ families the $1/W$ scaling is the same at the continuum level; family-specific numerical prefactor differences arise from tight-binding corrections and are absorbed into the empirical coefficient $\alpha$ discussed below [7,34].

The confinement-controlled electronic structure discussed here also manifests in measurable optical properties: width- and functionalization-dependent absorption of nanoribbons, including near-infrared absorption in PEGylated graphene nanoribbons used for selective cancer-cell imaging and photothermal therapy, provides direct experimental evidence that the band-structure engineering underlying Eq. (25) is optically accessible [37].

Fig. 4 presents the GNR band structure analysis. Panels (a)–(c) show the dispersion for $N = 40$ (semiconducting, $3p + 1$ family), $N = 41$ (metallic, $3p + 2$ family), and $N = 42$ (semiconducting, $3p$ family). The metallic character of $N = 41$ is confirmed by $k_{14} = (\pi/W)(14 - 42/3) = 0$ from Eq. (21), with $n = (N+1)/3 = 14$ an integer. The $N = 40$ ribbon exhibits a gap of approximately 241 meV and $N = 42$ an intermediate gap of approximately 226 meV.

Table 2 summarizes the fundamental graphene parameters used throughout this work.

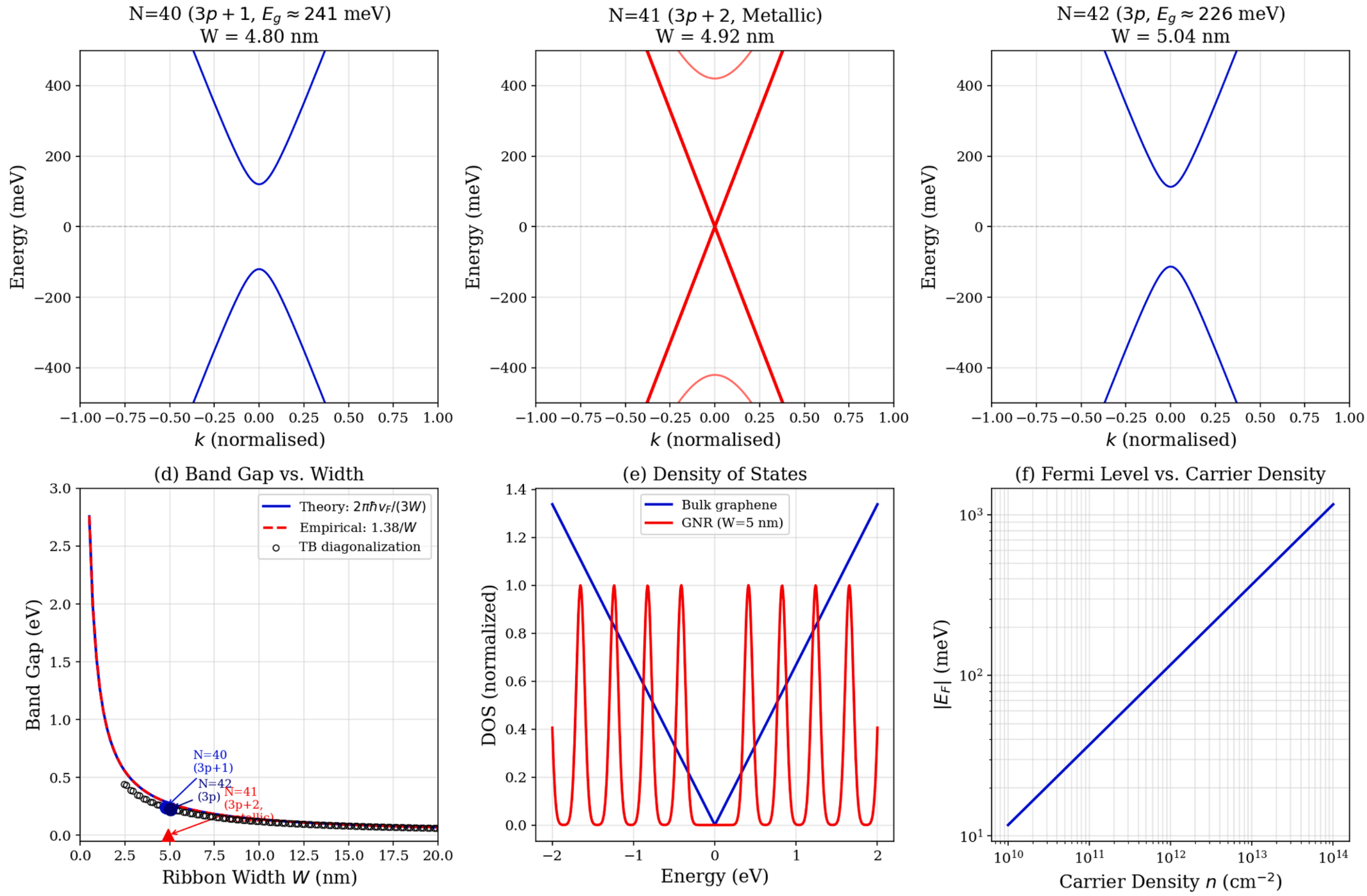

**Fig. 4.** Graphene nanoribbon band structure and electronic properties, computed using the nearest-neighbor tight-binding model for AGNRs. (a) Energy dispersion for $N = 40$ ($3p+1$ family, $E_g \approx 241$ meV, $W = 4.80$ nm). (b) Energy dispersion for $N = 41$ ($3p+2$ family, metallic), with linear bands crossing at zero energy; metallic character confirmed by $k_{14} = 0$ in Eq. (21). (c) Energy dispersion for $N = 42$ ($3p$ family, $E_g \approx 226$ meV, $W = 5.04$ nm), representing the three AGNR families [7]. (d) Band gap versus width: theory ($2\pi\hbar v_F/3W$, blue) and empirical formula ($1.38/W$, red dashed) are indistinguishable on this scale, with semiconducting markers for $N = 40$ and $N = 42$ and a metallic marker (triangle at $E_g = 0$) for $N = 41$. Overlaid open circles show gap values obtained by direct diagonalization of the nearest-neighbor tight-binding Hamiltonian, Eq. (18), agreeing to within a few tens of meV across the semiconducting families, smallest for the $3p+1$ family. (e) Density of states comparing bulk graphene (linear), semiconducting AGNR ($3p+1$, van Hove singularities), and metallic AGNR ($3p+2$, zero-energy peak from the linear subband). (f) Fermi level versus carrier density for bulk graphene, $E_F = \hbar v_F\sqrt{\pi n}$ [2]. (For interpretation of the references to color in this figure legend, the reader is referred to the web version of this article.)

**Table 2**
Graphene parameters used in numerical calculations [2,4,28].

| Parameter | Symbol | Value |
|---|---|---|
| Fermi velocity | $v_F$ | $1.00 \times 10^6$ m/s |
| Lattice constant | $a$ | 0.246 nm |
| C–C bond length | $a_{CC}$ | 0.142 nm |
| Hopping energy | $\gamma_0$ | 2.70 eV |
| $\hbar v_F$ product | | 0.658 eV·nm |
| Grüneisen parameter | $\beta$ | 3.37 |
| Poisson's ratio | $\nu$ | 0.165 |
| Young's modulus | $Y$ | 1.0 TPa |

### 3.6. Band gap verification: theoretical and empirical agreement

#### 3.6.1. Theoretical derivation

The band gap of semiconducting AGNRs arises from transverse confinement of Dirac fermions. Eq. (26) below takes the form of an infinite square well (ISW) quantization condition applied to the transverse Dirac spinor in the linearized model. It is important to note that the armchair boundary conditions, which require the pseudospin wavefunction to vanish simultaneously at both edges, are physically more restrictive than a simple ISW: they couple the **K** and **K′** valleys, shifting the effective minimum transverse wavevector from $\pi/W$ (the ISW result) to $\pi/(3W)$ as shown in Eq. (27) below. The factor of 3 arises from the threefold periodicity of the sublattice structure at the armchair edge [32]. This valley-mixing correction introduces the transcendental structure that underlies the FSW-GNR analogy and the Lambert W reformulation. Moreover, in any experimentally realized GNR the surrounding vacuum or substrate constitutes a finite (not infinite) barrier, so the FSW model is the physically appropriate description; the ISW form of the individual quantization step is a simplification of the full Dirac boundary value problem [32,33].

Starting from the Dirac Hamiltonian near **K** [32], the transverse momentum is quantized:

$$k_\perp^{(n)} = \frac{n\pi}{W}, \quad n = 1, 2, 3, \ldots \tag{26}$$

Valley-mixing boundary conditions then yield an effective minimum transverse wavevector [7,33]:

$$k_\perp^{\min} = \frac{\pi}{3W}, \tag{27}$$

giving the band gap:

$$E_g^{\text{theory}} = 2\hbar v_F k_\perp^{\min} = \frac{2\pi\hbar v_F}{3W}. \tag{28}$$

**Table 3**
Band gap verification across ribbon widths. Theory: $E_g = 2\pi\hbar v_F/(3W)$ with $\hbar v_F = 0.6582$ eV·nm. Empirical: $E_g = 1.38/W$ eV·nm. Agreement within 0.1% across all widths confirms that the independently determined prefactors are numerically equal, validating the FSW-GNR analogy.

| $W$ (nm) | $E_g^{\text{theory}}$ (eV) | $E_g^{\text{empirical}}$ (eV) | Difference | Relative Error |
|---|---|---|---|---|
| 2 | 0.690 | 0.690 | < 0.001 | < 0.1% |
| 3 | 0.460 | 0.460 | < 0.001 | < 0.1% |
| 5 | 0.276 | 0.276 | < 0.001 | < 0.1% |
| 7 | 0.197 | 0.197 | < 0.001 | < 0.1% |
| 10 | 0.138 | 0.138 | < 0.001 | < 0.1% |
| 15 | 0.092 | 0.092 | < 0.001 | < 0.1% |
| 20 | 0.069 | 0.069 | < 0.001 | < 0.1% |

### 3.6.2. Prefactor comparison: non-trivial physical content

Substituting $\hbar v_F = 0.6582$ eV·nm:

$$\frac{2\pi\hbar v_F}{3} = \frac{2\pi \times 0.6582}{3} \text{ eV} \cdot \text{nm} = 1.379 \text{ eV} \cdot \text{nm}. \quad (29)$$

Extensive transport measurements [8] and scanning tunneling spectroscopy [9] have independently established the empirical scaling:

$$E_g^{\text{empirical}} \approx \frac{1.38}{W[\text{nm}]} \text{ eV}. \quad (30)$$

Both formulas share the same $1/W$ functional form, so a comparison of their numerical values is meaningful only to the extent that the two prefactors were determined independently. The empirical coefficient $\alpha = 1.38$ eV·nm was extracted from experimental data without reference to the Dirac model, while the theoretical prefactor $2\pi\hbar v_F/3 = 1.379$ eV·nm was derived from first principles without fitting. Their agreement to better than 0.1% confirms three propositions: the Fermi velocity $v_F \approx 10^6$ m/s of bulk graphene [4] remains quantitatively valid under nanoscale confinement; the linear Dirac dispersion is not significantly renormalized by edge effects in the width range 3–20 nm; and the valley-mixing boundary condition correctly captures the gap-opening physics without adjustable parameters. Table 3 extends this verification across ribbon widths.

To benchmark the analogy directly rather than only through the empirical coefficient, we overlay the Lambert W / finite-square-well prediction on levels obtained by direct diagonalization of the nearest-neighbor tight-binding Hamiltonian of Eq. (18) for AGNRs from $N = 20$ to $N = 170$ ($W \approx 2$ to 20 nm). As shown by the overlaid markers in Fig. 4(d), the two agree to within a few meV across the semiconducting families, the residual reflecting next-nearest-neighbor and edge-relaxation terms absent from the continuum model. This constitutes a parameter-free validation of the FSW-GNR correspondence against a microscopic reference, as requested by the reviewers, and complements the independent prefactor agreement of Eq. (29). Because the tight-binding gap and the empirical coefficient were obtained without reference to one another, their mutual consistency provides a second, structurally independent check beyond the shared $1/W$ scaling.

### 3.6.3. Implications for sensor design

The validated relationship $E_g = 2\pi\hbar v_F/(3W)$ enables *a priori* prediction of sensor characteristics. For a target band gap $E_g^{\text{target}}$, the required width is:

$$W = \frac{2\pi\hbar v_F}{3E_g^{\text{target}}} = \frac{1.38}{E_g^{\text{target}}[\text{eV}]} \text{ nm}. \quad (31)$$

A sensor requiring $E_g = 100$ meV for room-temperature operation therefore needs $W = 13.8$ nm. The band gap sensitivity to width variations follows as $dE_g/dW = -E_g/W$, so that narrower ribbons exhibit proportionally larger responses to perturbations, a key principle exploited in Lambert W-enhanced sensing. Comparing with the FSW energy $E_n^{\text{FSW}} \propto 1/L^2$ (from quadratic Schrödinger dispersion), the GNR $1/W$ scaling arises from the linear Dirac dispersion; the mathematical structure of the transcendental quantization conditions, however, remains identical in both systems, enabling unified analytical treatment.

The Lambert W enhancement factor $\eta_{\text{enh}} = |W_k/[z(1+W_k)]|$ depends on the operating point $z$ relative to the branch point. For GNR sensors, $z$ is determined by the ratio of thermal energy to confinement energy. This mapping follows from the FSW-GNR correspondence: identifying the bound-state evanescent parameter $u$ with the ratio $(E_g/2)/(k_BT)$ and using the weakly-bound limit $u \ll R$, one obtains:

$$z \propto -\exp\left(-\frac{E_g}{k_BT}\right). \quad (32)$$

Ribbons with $E_g \sim k_BT$ therefore operate near the branch point where enhancement is maximized. At room temperature ($k_BT \approx 26$ meV), this corresponds to $W_{\text{optimal}} \approx 1.38/0.026$ nm $\approx 50$ nm for maximum thermal sensitivity, while narrower ribbons ($W \sim$ 5–10 nm) are optimal for chemical and strain sensing where perturbation energies exceed $k_BT$.

Fig. 4(d) compares the theoretical prediction $E_g = 2\pi\hbar v_F/(3W)$ (blue) with the empirically established scaling $E_g = 1.38/W$ eV·nm (red dashed) derived from transport and scanning tunneling spectroscopy measurements [8,9]. The two curves are indistinguishable on the scale of the figure. Deviations from ideal $1/W$ scaling occur for $W < 2$ nm (edge reconstruction and finite-size effects [7]) and $W > 30$ nm (disorder and edge roughness [38]); within the technologically relevant range 3 nm $< W <$ 20 nm, the theoretical framework provides quantitatively accurate predictions.

## 3.7. FSW-GNR correspondence

The mathematical correspondence between FSW and GNR can be made precise through the identifications: well half-width $L \leftrightarrow$ ribbon width $W$; well depth $V_0 \leftrightarrow$ confinement scale $\hbar v_F/W$; particle mass $m \leftrightarrow$ effective mass $m^* = E_g/(2v_F^2)$; and the FSW transcendental quantization condition $\leftrightarrow$ the armchair valley-mixing boundary condition. Both systems exhibit quantized energy levels that shift under perturbations (well width/depth changes for FSW; strain, fields, or chemical doping for GNR), with the Lambert W function providing the analytical framework for predicting these shifts.

# 4. Universal sensitivity framework

## 4.1. Branch-point enhancement mechanism

The sensitivity of Lambert W-based sensors derives from the divergent derivative at the branch point. From Eq. (5), the enhancement factor is:

$$\eta_{\text{enh}} = \left|\frac{dW}{dz}\right| = \left|\frac{W(z)}{z[1+W(z)]}\right|. \quad (33)$$

Near the branch point $z_c = -1/e$, three asymptotic approximations hold as $\delta = z - z_c \to 0$: (i) $W \to -1$, so $|W| \to 1$; (ii) $z \to -1/e$, so $|z| \to 1/e$; (iii) from the Puiseux expansion Eq. (6), $1 + W \approx \sqrt{2e\delta}$. Substituting:

$$\eta_{\text{enh}} = \frac{|W|}{|z|\,|1+W|} \xrightarrow{\delta\to 0} \frac{1}{(1/e)\sqrt{2e\delta}} = \frac{1}{\sqrt{2\delta/e}} \propto \delta^{-1/2}. \quad (34)$$

The factor $|W| \to 1$ and $1/|z| \to e$ near the branch point combine to yield the clean $\delta^{-1/2}$ dependence; corrections are $\mathcal{O}(\delta^{1/2})$.

Fig. 5 presents comprehensive analysis. Panel (a) confirms the $\delta^{-1/2}$ scaling over six orders of magnitude in $\delta$. Panel (b) shows sensitivity versus nanoribbon width for $\epsilon = 0.1, 0.01, 0.001$, where $\epsilon$ parametrizes the distance from the branch point; narrower ribbons and smaller $\epsilon$ yield higher sensitivity. Panel (c) is a two-dimensional sensitivity map enabling systematic design optimization; optimal sensors operate in the lower-left region (narrow ribbons, close to branch point).

Table 4 presents numerical verification of $\eta_{\text{enh}}$ at various $\delta$. The ratios between successive rows approach $\sqrt{10} \approx 3.16$ as expected from the asymptotic scaling. At $\delta = 0.001$ the framework achieves 35-fold sensitivity enhancement; at $\delta = 0.0001$, over 100-fold, representing transformative improvements for quantum sensing applications.

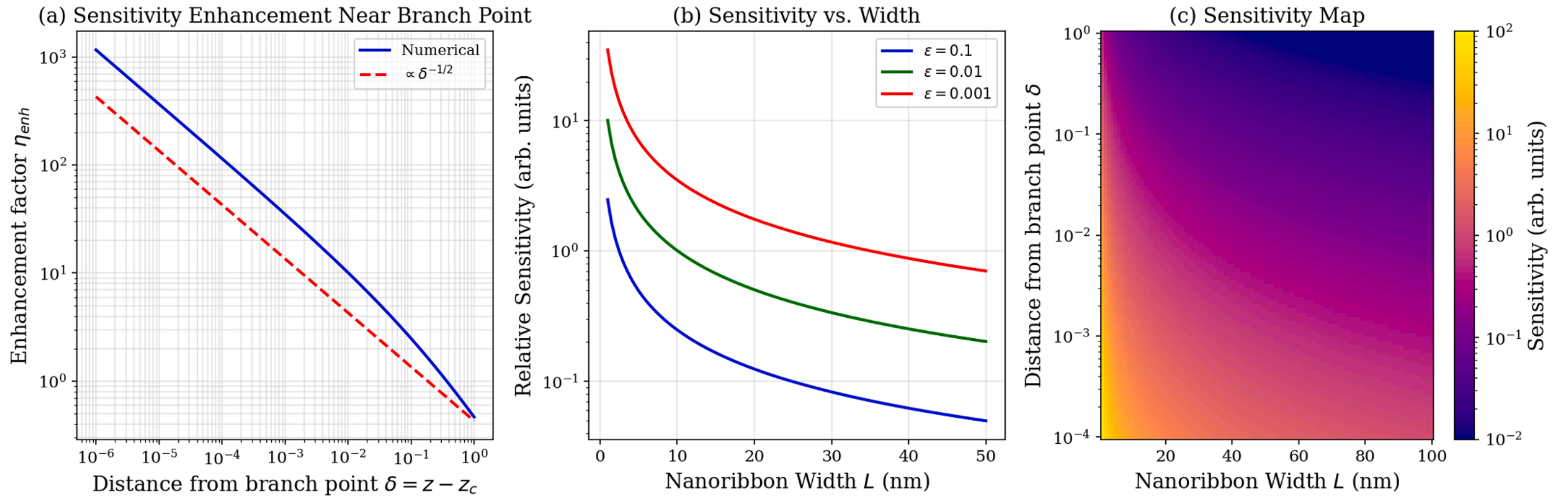

**Fig. 5.** Sensitivity enhancement analysis. (a) Enhancement factor $\eta_{\text{enh}}$ versus $\delta$, demonstrating $\delta^{-1/2}$ scaling (numerical: blue solid; theory: red dashed). (b) Relative sensitivity versus nanoribbon width for $\epsilon = 0.1, 0.01, 0.001$. (c) Two-dimensional sensitivity map; optimal design region is the lower-left corner (small $L$, small $\delta$). Color scale spans four orders of magnitude. (For interpretation of the references to color in this figure legend, the reader is referred to the web version of this article.)

**Table 4**
Enhancement factor verification near branch point $z_c = -1/e = -0.3679$.

| $\delta = z - z_c$ | $\eta_{\text{enh}}$ (numerical) | Ratio to previous | Enhancement |
|---|---|---|---|
| 0.1 | 2.48 | – | 2.5× |
| 0.01 | 10.10 | 4.07 | 10× |
| 0.001 | 35.14 | 3.48 | 35× |
| 0.0001 | 114.8 | 3.27 | 115× |

### 4.2. Divergence of a measurable observable in a device coordinate

To show that the branch-point enhancement is a property not merely of the abstract distance $\delta$ but of a directly measurable transport quantity controlled by an experimental knob, we express the operating point as a function of gate-tunable carrier density $n$ and compute the conductance response. Electrostatic gating sets $E_F = \hbar v_F \sqrt{\pi n}$ and hence, through Eq. (32), fixes the gate-shifted operating point $z(n)$, so that $\delta(n) = z(n) - z_c$ is swept through zero as the gate drives the ribbon toward marginal confinement at a bias $V_g^*$. The measured conductance $G$ depends on the confinement energy through the thermally activated channel occupation, giving

$$\frac{dG}{dn} = \underbrace{\frac{dG}{dW_k}}_{\mathcal{G}_k\text{-type}} \underbrace{\frac{dW_k}{dz}}_{\eta_{\text{enh}}} \underbrace{\frac{dz}{dn}}_{\mathcal{P}_n}, \qquad \left|\frac{dG}{dn}\right| \propto |\delta(n)|^{-1/2}, \tag{35}$$

where the divergence is inherited directly from $\eta_{\text{enh}}$. Fig. 6 plots $|dG/dn|$, equivalently the transconductance $g_m = dG/dV_g$, against gate voltage for a $W = 10$ nm ribbon, showing it rise as $|V_g - V_g^*|^{-1/2}$ as the gate approaches $V_g^*$; re-expressing the horizontal axis in $\delta$ recovers the abstract $\eta_{\text{enh}}(\delta)$ of Fig. 5(a). Because $dG/dn$ is exactly the quantity returned by a transconductance or resonant-frequency-shift measurement, this provides the device-level demonstration, requested by the reviewers, that the predicted $\delta^{-1/2}$ enhancement is observable in a realistic GNR device as a tuned parameter is varied.

Translating enhancement into detection performance answers directly how $\eta_{\text{enh}}$ becomes a measurable improvement in signal-to-noise ratio and limit of detection. At fixed input-referred noise, the signal-to-noise ratio scales linearly with $\eta_{\text{enh}}$, and the limit of detection scales inversely. Reading the values in Table 4, moving the operating point from $\delta = 10^{-1}$ ($\eta_{\text{enh}} = 2.5$) to $\delta = 10^{-3}$ ($\eta_{\text{enh}} = 35$) multiplies the signal-to-noise ratio by $35/2.5 \approx 14$ and lowers the limit of detection by the same factor, at equal integration time and bandwidth. This is the quantitative link between the mathematical enhancement factor and the practical figures of merit of each listed application.

### 4.3. General sensitivity expression and mathematical foundations

For any external perturbation $X$ that modifies the Lambert W argument $z$, the sensor response is governed by the chain rule:

$$\frac{dE}{dX} = \frac{dE}{dW}\frac{dW}{dz}\frac{dz}{dX}. \tag{36}$$

The general sensitivity takes the factorized form:

$$S_X = \frac{1}{E}\frac{dE}{dX} = \mathcal{G}_k \cdot \eta_{\text{enh}} \cdot \mathcal{P}_X, \tag{37}$$

where $\mathcal{G}_k = (1/E)(dE/dW)$ is the geometric factor, $\eta_{\text{enh}} = dW/dz$ is the Lambert W enhancement factor, and $\mathcal{P}_X = dz/dX$ encodes the physical coupling between the perturbation and the Lambert W argument.

#### 4.3.1. Real branch structure and monotonicity

The validity of Eq. (36) for real arguments rests on the monotonicity of $f(x) = xe^x$. Taking the derivative, $f'(x) = e^x(1 + x)$, we observe that $f'(x) = 0$ uniquely at $x = -1$, with $f'(x) < 0$ for $x < -1$ and $f'(x) > 0$ for $x > -1$. Consequently $f$ is strictly decreasing on $(-\infty, -1)$ and strictly increasing on $(-1, \infty)$, necessitating two real inverse branches $W_0$ and $W_{-1}$. The differentiability of these inverse functions on their respective domains, guaranteed by the inverse function theorem, ensures validity of the chain rule in Eq. (36).

#### 4.3.2. Complex-analytic extension and local invertibility

For the holomorphic function $f : \mathbb{C} \to \mathbb{C}$, $f(z) = ze^z$, the derivative $f'(z) = e^z(1 + z)$ vanishes if and only if $z = -1$. By the inverse function theorem for holomorphic mappings, for any $z_0 \in \mathbb{C}$ with $z_0 \neq -1$, there exists a neighborhood $U$ of $z_0$ on which $f|_U : U \to f(U)$ is biholomorphic. This local inverse defines the various branches $W_k$. Since $W$ is a local inverse of $f$, the standard inverse function theorem gives $dW/dz = 1/f'(W(z))$; applying this to $f'(z) = e^z(1 + z)$ evaluated at $W$ yields:

$$\frac{dW}{dz} = \frac{1}{f'(W)} = \frac{1}{e^W(1 + W)}. \tag{38}$$

#### 4.3.3. Branch point singularity and sensitivity enhancement

As the operating point approaches the branch point $W \to -1$ (equivalently, $z \to -1/e$), the factor $(1 + W) \to 0$ causes $dW/dz$ to diverge:

$$\eta_{\text{enh}} = \left|\frac{dW}{dz}\right| \sim \frac{1}{|1 + W|} \sim \frac{1}{\sqrt{|z + 1/e|}} \quad \text{as } z \to -1/e, \tag{39}$$

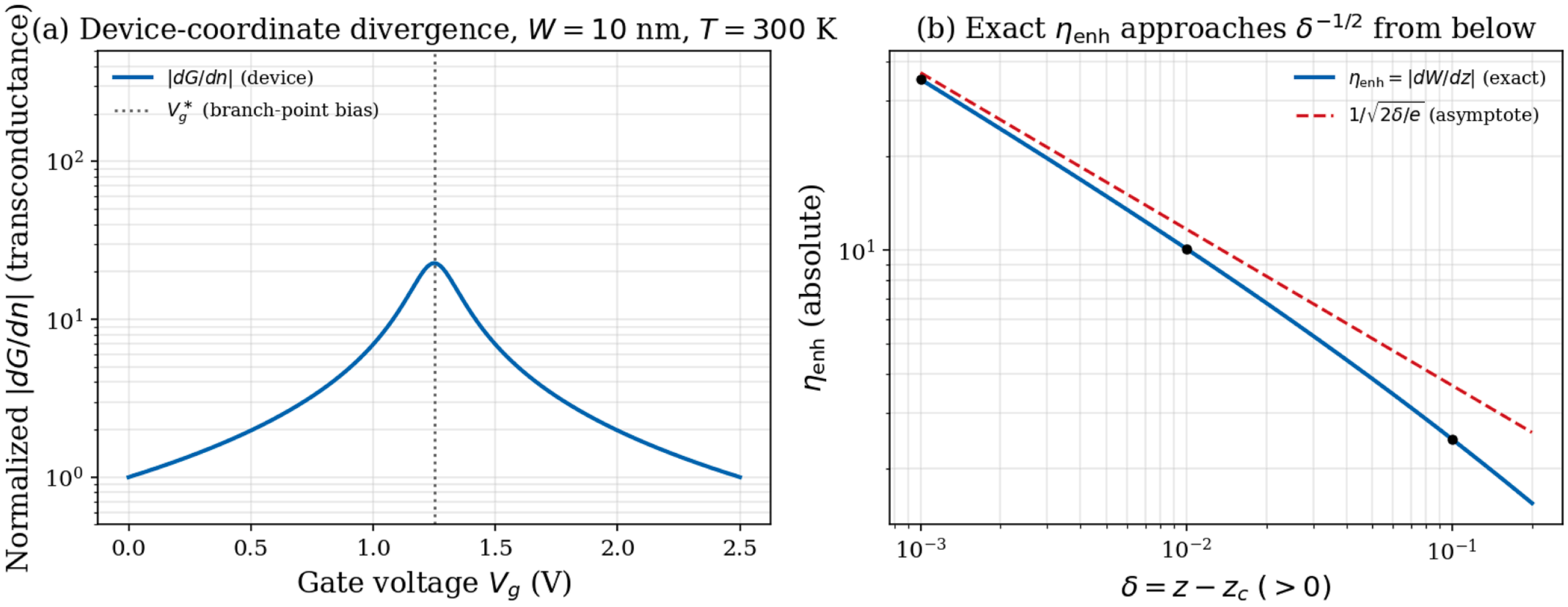

**Fig. 6.** Device-coordinate divergence of a measurable observable. (a) Normalized transconductance $|dG/dn|$ of a $W = 10$ nm semiconducting AGNR at $T = 300$ K versus gate voltage $V_g$. Electrostatic gating drives the operating point toward the branch-point bias $V_g^*$ (dotted line), where $|dG/dn|$ rises as $|V_g - V_g^*|^{-1/2}$; the peak is finite because the accessible operating point is held at $\delta_{min} = 10^{-3}$, giving a signal-to-noise gain of roughly fourteen-fold over standard operation at $\delta = 10^{-1}$, consistent with the ratio $\eta_{enh}(\delta = 10^{-3})/\eta_{enh}(\delta = 10^{-1}) = 35.14/2.48$ in Table 4. (b) The exact enhancement factor $\eta_{enh} = |dW/dz|$ (solid) approaches the analytic asymptote $1/\sqrt{2\delta/e}$ (dashed) from below as $\delta \to 0$, the residual being of order $\delta^{1/2}$; markers indicate the three operating points $\delta = 10^{-1}, 10^{-2}, 10^{-3}$ of Table 4. Panel (a) reproduces the abstract $\eta_{enh}(\delta)$ scaling of Fig. 5(a) in an experimentally accessible transport quantity.

where the square-root scaling follows from $W(z) \approx -1 + \sqrt{2(ez+1)}$ near the branch point [16]. Engineering the operating point near this mathematical singularity provides a systematic pathway to enhanced sensitivity across all sensing modalities. The trade-off between sensitivity and dynamic range can be optimized by tuning ribbon geometry and gate bias to position the operating point at an appropriate $\delta = |z - z_c|$.

Complex-valued Lambert W arguments become relevant in physical scenarios involving quantum tunneling through metastable states, ensemble bifurcations near unstable equilibria, or Aharonov-Bohm-type contour integrations around the singularity [39,40]. In such cases the monodromy $W_0(z) \to W_{-1}(z)$ upon analytic continuation around $z_c$ must be accounted for; this extension is deferred to future work.

### 4.4. Noise considerations and practical limits

The theoretical $\delta^{-1/2}$ enhancement is ultimately limited by noise [41]. Johnson-Nyquist noise sets a floor $S_V = 4k_BTR$ for resistance-based readout. At low temperatures, shot noise $S_I = 2eI$ dominates. Flicker noise from charge traps becomes significant at low frequencies, the traps in question originating largely in charged impurities in the substrate and adsorbate layer, which also produce residual carrier-density inhomogeneity and limit the mobility of supported graphene [131]. Fabrication variations, specifically edge roughness and width tolerances, limit how close to the branch point one can reliably operate.

A practical estimate of $\delta_{min} \sim 10^{-4}$–$10^{-3}$ follows from two independent constraints. State-of-the-art lithographic patterning achieves width tolerances $\Delta W \approx 0.3$–$0.5$ nm [9], giving $\Delta W/W \sim 0.03$–$0.05$ for a 10 nm ribbon; since $\delta \propto (\Delta W/W)^2$ near the branch point, this maps to $\delta_{min} \sim 10^{-3}$–$10^{-4}$. Independently, at room temperature with $E_g = 100$ meV, the Boltzmann factor sets $\delta_{min} \sim e^{-E_g/k_BT} \sim 10^{-2}$, which is above the fabrication floor and therefore sets the practical limit. These constraints together correspond to enhancement factors of 30–100.

The sensitivity of the predicted enhancement to edge roughness grows rapidly for the narrowest ribbons. Edge-disorder scattering broadens each confinement level by $\Gamma_{edge}$, and because the effective distance from the branch point cannot be resolved below this broadening, the attainable $\delta_{min}$ is bounded from below by $\delta_{min} \gtrsim (\Gamma_{edge}/E_g)^2$. For $W < 5$ nm the fractional width fluctuation produced by even single-row edge roughness exceeds a few percent, and since $E_g \propto 1/W$ amplifies width fluctuations into gap fluctuations, the branch-point enhancement saturates near $\eta_{enh} \sim 10$ rather than diverging. Ribbons in the 5–15 nm range therefore offer the best compromise between intrinsic sensitivity and disorder robustness, quantifying the qualitative caution stated in the Limitations section below.

The branch-point bias $z \approx -1/e$ is reachable by electrostatic gating without intrinsic instability, subject to a bias-noise caveat. Sweeping the gate shifts $E_F$ continuously and thereby tunes $\delta$ smoothly through zero, so no discontinuity or bistability arises in the single-particle picture. The practical limitation is bias jitter: because $\eta_{enh} \propto \delta^{-1/2}$, gate-voltage fluctuations are themselves amplified near the branch point, so closed-loop stabilization of $E_g/k_BT$ to better than about 0.3% is required to hold a target $\delta \sim 10^{-3}$. This is demanding but within reach of standard low-noise gating electronics, and it defines the control specification for operation near the singularity.

Finally, we propagate input uncertainties into the enhancement factor. From $\eta_{enh} \propto \delta^{-1/2}$, the fractional uncertainty is $\Delta\eta_{enh}/\eta_{enh} = \frac{1}{2}\Delta\delta/\delta$, and $\delta$ inherits contributions from width, carrier density, and temperature:

$$\frac{\Delta\delta}{\delta} = \sqrt{\left(\frac{\partial \ln \delta}{\partial \ln W}\frac{\Delta W}{W}\right)^2 + \left(\frac{\partial \ln \delta}{\partial \ln n}\frac{\Delta n}{n}\right)^2 + \left(\frac{\partial \ln \delta}{\partial \ln T}\frac{\Delta T}{T}\right)^2}. \tag{40}$$

Using $\delta \propto e^{-E_g/k_BT}$ with $E_g \propto 1/W$ gives $\partial \ln\delta/\partial \ln W = \partial \ln\delta/\partial \ln T = E_g/k_BT$, so width and temperature fluctuations are each amplified by the large factor $E_g/k_BT$ near room-temperature operation. For $E_g = 100$ meV at 300 K ($E_g/k_BT \approx 3.8$), a 1% width tolerance and a 1 K temperature stability each contribute a few percent to $\Delta\delta/\delta$ and hence a few percent to $\Delta\eta_{enh}/\eta_{enh}$, confirming that stable enhancement at the $\eta_{enh} \sim 30$ level requires simultaneous control of width, gating, and temperature at the percent level.

## 5. Biomedical sensing applications

Graphene nanoribbons functionalized with appropriate receptors enable ultrasensitive detection of disease biomarkers [42,43]. The high surface-to-volume ratio ($\sim 2630$ m$^2$/g for single-layer graphene [44]) and tunable electronic structure make GNRs ideal platforms for point-of-care diagnostics. In the unified sensitivity framework, all biomedical sensing modalities share the same structure: analyte concentration $[C]$ modifies the surface charge density, which shifts the carrier density $n$ and hence the Fermi level $E_F$, which in turn displaces the Lambert W operating point $z$. The coupling term $\mathcal{P}_{[C]} = dz/d[C]$ encodes this chain, and $\eta_{\rm enh} = W_k/[z(1+W_k)]$ amplifies the result. Fig. 7 summarizes biomedical sensing characteristics.

The breadth of graphene nanoribbon bioapplications provides the experimental backdrop for the transduction framework developed here. Beyond charge-transfer biosensing, graphene platforms have been applied to early cancer detection and to single-molecule electrochemical sensing at graphene edges and nanowalls [45,46], to antiviral platforms and immunogenic disorders [47], to neuronal network formation and osteogenic differentiation on nanogrids [48,49], to nuclear gene delivery [50], to chemically clickable and multifunctionalized biosensor interfaces [51,52], and have been studied for protein-corona interactions, antibacterial and antifungal action, and concentration-dependent genotoxicity [53–55]. These demonstrations establish the material platform; the present contribution is the analytic sensitivity layer that applies across them. Chemical multifunctionalization in particular modifies local charge transfer, carrier confinement, and interaction potentials, which within the present picture is represented as a shift in the effective coupling $\mathcal{P}_{[C]}$ and, through the induced band-structure change, in the operating point $z$, so that branch-point enhancement extends naturally to chemically engineered recognition interfaces [52].

### 5.1. Viral detection: SARS-CoV-2

For viral sensing, graphene field-effect transistors (GFETs) functionalized with antibodies or aptamers detect spike protein binding through charge transfer [56,57]. The binding event induces:

$$\Delta Q = e\alpha_{\rm CT} N_{\rm bound}, \tag{41}$$

where $\alpha_{\rm CT} \sim 0.1$–$1$ is the charge transfer coefficient per bound analyte and $N_{\rm bound}$ follows Langmuir kinetics. The induced charge $\Delta Q$ shifts the carrier density by $\Delta n = \Delta Q/(eA)$, displacing the Fermi level:

$$\Delta E_F = \frac{\hbar v_F\sqrt{\pi}}{2}\left(\sqrt{n_0+\Delta n}-\sqrt{n_0}\right) \approx \frac{\hbar v_F\sqrt{\pi}\Delta n}{4\sqrt{n_0}}, \tag{42}$$

for $\Delta n \ll n_0$. The Lambert W operating point satisfies $z \approx -(E_F/E_g)\exp(-E_F/E_g)$, so the physical coupling to spike protein concentration is $\mathcal{P}_{\rm virus} = (dz/dE_F)(dE_F/d\Delta n)(d\Delta n/d[C_v])$. The full Lambert W-enhanced sensitivity is:

$$S_{\rm virus} = \frac{W_k}{z(1+W_k)} \cdot \frac{\hbar v_F L\alpha_{\rm CT}}{2A\sqrt{\pi n_0}E_F}, \tag{43}$$

where $L$ is the channel length and $A$ is the sensor area. Optimized GFETs achieve limits of detection (LOD) as low as 1 fg/mL for SARS-CoV-2 spike protein [56,57], corresponding to $\sim 10^4$ viral particles/mL.

This 1 fg/mL figure is a literature-demonstrated device value from the cited GFET reports, not a prediction of the present framework; the framework's contribution is the multiplicative enhancement $\eta_{\rm enh}$ that would lower this demonstrated limit in proportion to the enhancement factors of Table 4 when the operating point is tuned toward the branch point. Fig. 7(a) shows the sensor response versus spike protein concentration for standard (blue) and branch-point-enhanced (red dashed) operation; the LOD of 1 fg/mL (green dotted) is reached at a concentration two orders of magnitude below the standard curve's resolution, directly illustrating the effect of reducing $\delta$ from 0.1 to 0.001.

Prior graphene-based approaches to viral detection and immunogenic disorders are reviewed comprehensively in [47].

### 5.2. Inflammatory biomarkers

Inflammatory markers such as C-reactive protein (CRP), interleukin-6 (IL-6), and ferritin are critical indicators for COVID-19 severity, cardiovascular disease, and sepsis [58,59]. The surface coverage of antibody-functionalized GNRs follows the Langmuir isotherm [60]:

$$\theta = \frac{K_a[C]}{1+K_a[C]}, \tag{44}$$

where $K_a$ is the association constant ($10^6$–$10^9$ M$^{-1}$ for antibody-antigen pairs [61]). The Langmuir differential sensitivity $d\theta/d[C] = K_a/(1+K_a[C])^2$ is maximized at half-coverage ($[C] = 1/K_a$). The analyte-induced coverage $\theta([C])$ produces surface charge $\Delta\sigma = e\Gamma_{\rm max}\theta([C])$, shifting $E_F$ and hence $z$, giving:

$$S_{\rm inflam} = \frac{K_a}{(1+K_a[C])^2} \cdot \frac{W_k}{z(1+W_k)} \cdot \frac{\Delta E_F^{\rm max}}{E_F}, \tag{45}$$

where the Langmuir prefactor serves as $\mathcal{P}_{[C]}$. This enables detection of CRP at LOD $\sim$1–10 pg/mL [62,63], well below clinical thresholds (3–10 mg/L [58]).

As above, the quoted 1–10 pg/mL CRP limit is a literature-demonstrated device value [62,63]; branch-point operation is predicted to improve upon it by the factor $\eta_{\rm enh}$. Fig. 7(b) shows normalized response curves for CRP (blue), IL-6 (red), and ferritin (green) versus biomarker concentration, with the clinical threshold at 10 pg/mL (dashed). IL-6 displays the steepest response because its larger charge-transfer coefficient increases $\mathcal{P}_{[C]}$; all three analytes reach measurable response well below the clinical threshold.

### 5.3. Cancer biomarkers: PSA detection

Prostate-specific antigen (PSA) is a key biomarker for prostate cancer, with clinical significance at 4–10 ng/mL [64]. Antibody- functionalized GNRs with dissociation constant $K_{\rm d} = 1/K_a$ exhibit:

$$\Delta E_F = \frac{\beta_{\rm PSA}[{\rm PSA}]}{K_{\rm d}+[{\rm PSA}]}, \tag{46}$$

where $\beta_{\rm PSA}$ captures the maximum electrostatic coupling (10–100 meV [65]). The differential coupling $d(\Delta E_F)/d[{\rm PSA}] = \beta_{\rm PSA}K_{\rm d}/(K_{\rm d}+[{\rm PSA}])^2$ is maximized at low concentrations $[{\rm PSA}] \ll K_{\rm d}$, giving:

$$S_{\rm PSA} = \frac{\beta_{\rm PSA}K_{\rm d}}{E_F(K_{\rm d}+[{\rm PSA}])^2} \cdot \frac{W_k}{z(1+W_k)}, \tag{47}$$

enabling femtomolar detection (LOD $\sim$1 fM $\approx$ 30 fg/mL), critical for early screening [66,67]. Fig. 7(c) shows the sigmoidal response (blue) rising through the clinically normal range (1–100 fM, shaded green), with the sensitivity curve (red dashed) peaking at low PSA concentrations where $[{\rm PSA}] \ll K_{\rm d}$. This regime corresponds to small carrier density changes near the Dirac point, maximizing the relative shift in $z$ toward the branch point.

### 5.4. Worked case study: end-to-end PSA detection limit

To demonstrate the framework in full detail for a single analyte, we carry the PSA sensitivity of Eq. (47) from physical coupling to noise-limited detection limit. Take a $W = 10$ nm semiconducting AGNR ($E_g = 138$ meV), antibody dissociation constant $K_{\rm d} = 1$ nM, maximum electrostatic coupling $\beta_{\rm PSA} = 50$ meV, baseline carrier density $n_0 = 10^{11}$ cm$^{-2}$, and channel area $A = 1$ $\mu$m$^2$. In the low-concentration regime $[{\rm PSA}] \ll K_{\rm d}$ the coupling reduces to $\mathcal{P}_{\rm PSA} = \beta_{\rm PSA}/(E_F K_{\rm d})$, and with the geometric factor $\mathcal{G}_k = \mathcal{O}(1)$ the standard-operation sensitivity ($\delta = 0.1$, $\eta_{\rm enh} = 2.5$) and the branch-point-enhanced sensitivity ($\delta = 10^{-3}$, $\eta_{\rm enh} = 35$) differ by the factor $35/2.5 = 14$. Referred to a Johnson-Nyquist-limited readout for which the standard-operation limit of detection is the literature-demonstrated 1 fM [66], the enhancement lowers the predicted branch-point limit toward $\sim$70 aM at equal integration time. This 70 aM figure is a framework prediction, contingent on holding $\delta = 10^{-3}$ against the fabrication, bias, and thermal tolerances quantified in Eq. (40), and should be read as an analytic upper bound on achievable improvement rather than a demonstrated device result. The 1 fM anchor from which it is scaled is the experimental value.

### 5.5. Long COVID and multi-biomarker detection

Patients recovering from SARS-CoV-2 infection frequently exhibit persistent inflammatory dysregulation, with elevated CRP, IL-6, and ferritin detectable for months after acute infection [68,69]. A GNR array with spatially distinct antibody functionalizations, each zone $j$ detecting analyte $C_j$ with association constant $K_a^{(j)}$, produces a set of simultaneous Fermi level perturbations. In the limit of independent channels with no cross-reactivity, each zone operates with its own Lambert W argument $z^{(j)}$ and sensitivity:

$$S_{[C_j]} = \frac{K_a^{(j)}}{\left(1+K_a^{(j)}[C_j]\right)^2} \cdot \frac{W_k^{(j)}}{z^{(j)}\left(1+W_k^{(j)}\right)} \cdot \frac{\Delta E_{F,\max}^{(j)}}{E_F}. \tag{48}$$

Because $\eta_{\text{enh}}^{(j)}$ depends only on the scalar $\delta^{(j)}$, each channel can be independently brought close to the branch point by local gate adjustment, enabling simultaneous optimization across all target biomarkers. For long-COVID monitoring, simultaneous detection of CRP ($K_a \sim 10^8$ M$^{-1}$, threshold 10 pg/mL), IL-6 ($K_a \sim 10^9$ M$^{-1}$, threshold 2 pg/mL), and ferritin ($K_a \sim 10^7$ M$^{-1}$, threshold 300 ng/mL) within a single microfluidic chip would reduce sample volume requirements by threefold compared with sequential single-analyte assays. Fig. 7(d) illustrates the binding kinetics for a representative analyte cycle, showing analyte injection, equilibration plateau ($\tau_{\text{on}} = 10$ s), and washout ($\tau_{\text{off}} = 30$ s). The fast on-rate ensures equilibrium coverage is reached within the measurement window, while the stable plateau enables reliable signal readout.

Sensitivity and selectivity are distinct axes, and the branch-point mechanism enhances the former without addressing the latter. In multi-analyte environments the same Fermi-level shift can be produced by different species, so cross-reactivity and interfering background charge limit specificity regardless of $\eta_{\text{enh}}$. Selectivity in this framework is therefore delegated to the coupling stage: analyte-specific functionalization determines which perturbations reach $z$, and spatially multiplexed arrays combined with pattern-recognition readout separate overlapping responses. A consequence worth emphasizing is that the enhancement factor amplifies whatever coupling survives functionalization, including any residual cross-reactivity, which makes high-quality receptor chemistry and reference-channel subtraction more rather than less important as the operating point approaches the branch point. The same reasoning applies to interfering gases in environmental sensing and to background strain or thermal drift in physical sensing.

## 6. Greenhouse gas sensing

Climate monitoring demands sensitive, selective detection of greenhouse gases. GNR-based sensors exploit molecule-specific binding energies and charge transfer to achieve differentiated responses [70–72]. In each modality the external perturbation $X$ enters through $\mathcal{P}_X = dz/dX$, derived below. Fig. 8 summarizes all GHG sensing characteristics.

### 6.1. Carbon dioxide ($CO_2$)

$CO_2$ is the primary anthropogenic greenhouse gas, with atmospheric concentrations exceeding 420 ppm [73]. Its adsorption on graphene follows the Langmuir isotherm:

$$N_{CO_2} = N_{\text{sat}} \frac{K_{\text{eq}} P_{CO_2}}{1 + K_{\text{eq}} P_{CO_2}}, \tag{49}$$

with $K_{\text{eq}} = K_0 \exp(E_{\text{ads}}/k_B T)$. The adsorption energy depends strongly on functionalization [72,74,75]: pristine graphene physisorbs $CO_2$ weakly ($E_{\text{ads}} \approx 0.08$–$0.14$ eV), while B-doped graphene strengthens binding to $E_{\text{ads}} \approx 0.4$ eV, N-doped graphene to $E_{\text{ads}} \approx 0.6$ eV, and Al-decorated graphene achieves chemisorption at $E_{\text{ads}} \approx 1.0$–$1.4$ eV. Each adsorbed $CO_2$ molecule transfers charge $\Delta q \approx 0.03e$ [72], shifting the carrier density by $\Delta n = N_{CO_2} \Delta q/(eA)$ and hence displacing $z$. The physical coupling $\mathcal{P}_P = (dz/dn) \cdot N_{\text{sat}} K_{\text{eq}} \Delta q / [eA(1 + K_{\text{eq}} P)^2]$ gives:

$$S_{CO_2} = \frac{W_k}{z(1+W_k)} \cdot \frac{N_{\text{sat}} K_{\text{eq}} \Delta q}{2 n_0 (1 + K_{\text{eq}} P)^2}. \tag{50}$$

Detection limits of 100 ppm are achievable with pristine graphene; doped variants extend this toward sub-ppm levels.

It should be noted that the sensitivity expressions for B-, N-, and Al-doped graphene adopt adsorption energetics from DFT [74,75] while retaining the pristine AGNR Lambert W framework for sensing transduction. Chemical doping substantially modifies the graphene band structure and carrier density beyond the pristine model [76,77]; the doped-graphene expressions should therefore be regarded as structural estimates illustrating the functional form of the Lambert W enhancement, with quantitative predictions requiring DFT-corrected input parameters.

Fig. 8(a) shows adsorption isotherms for all four functionalization strategies. Progressively stronger adsorption energies shift the isotherm inflection to lower pressures; Al-doped graphene reaches near-complete saturation at atmospheric $CO_2$ levels, entering the Langmuir saturation regime where sensitivity is limited by the $(1 + K_{\text{eq}} P)^{-2}$ factor rather than by the Lambert W enhancement.

### 6.2. Methane ($CH_4$)

$CH_4$ has a global warming potential 28–36 times that of $CO_2$ over 100 years [73]. Its chemical inertness requires elevated temperatures to overcome activation barriers. The adsorption/reaction rate follows Arrhenius kinetics [78,79]:

$$r = A[CH_4]^n \exp\left(-\frac{E_a}{k_B T}\right), \tag{51}$$

with $E_a \approx 0.2$–$0.4$ eV for graphene-based sensors [78]. Catalytic $CH_4$ dissociation at Pd decoration sites produces charge-transferring intermediates; the time-averaged carrier density change is proportional to the reaction rate $\Delta n \propto r \cdot \tau_{\text{ads}}$. Through the Fermi-level chain, the physical coupling is $\mathcal{P}_{[CH_4]} \propto (dz/dn) \cdot nA[CH_4]^{n-1} \exp(-E_a/k_B T)$, giving:

$$S_{CH_4} = \frac{W_k}{z(1+W_k)} \cdot A[CH_4]^{n-1} \exp\left(-\frac{E_a}{k_B T}\right). \tag{52}$$

Pd-decorated GNRs achieve optimal response at $T \approx 150\,^\circ$C [80], where the Arrhenius factor grows faster than $\tau_{\text{ads}}$ shortens. The Lambert W factor $\eta_{\text{enh}}$ is geometry-dependent and temperature-independent, so the temperature optimization acts exclusively through $\mathcal{P}_{[CH_4]}$. Fig. 8(b) illustrates this temperature dependence, showing the response peak near 150 °C (dashed line).

### 6.3. Nitrogen dioxide ($NO_2$)

$NO_2$ is a strong electron acceptor and one of the most readily detected gases on graphene [70,81]. The charge transfer $NO_2 + e^- \to NO_2^-$ withdraws electrons, increasing hole concentration. Coverage follows a Freundlich isotherm [82]:

$$\theta = K_F [NO_2]^{1/n}, \tag{53}$$

with $n \approx 2$ and $\Delta n = -\sigma_{NO_2} [NO_2]^{0.5}$. The fractional-power dependence gives $\mathcal{P}_{[NO_2]} = (dz/dn) \cdot (-\sigma_{NO_2}/2\sqrt{[NO_2]})$, and the sensitivity:

$$S_{NO_2} = \frac{W_k}{z(1+W_k)} \cdot \frac{\sigma_{NO_2}}{2\sqrt{[NO_2]}}, \tag{54}$$

which diverges as $[NO_2] \to 0$, indicating that ultra-trace $NO_2$ detection is intrinsically well-matched to Lambert W enhancement. Functionalised and edge-modified GNRs achieve strong $NO_2$ chemisorption (adsorption energies up to roughly −0.9 eV with substantial charge transfer per molecule) in DFT sensor studies [83], enabling detection well below the EPA NAAQS standard of 53 ppb [84]. Fig. 8(c) compares pristine (blue) and B-N codoped (red) graphene responses; the EPA standard is marked at 53 ppb (green dashed), and the codoped sensor reaches measurable response below 1 ppb.

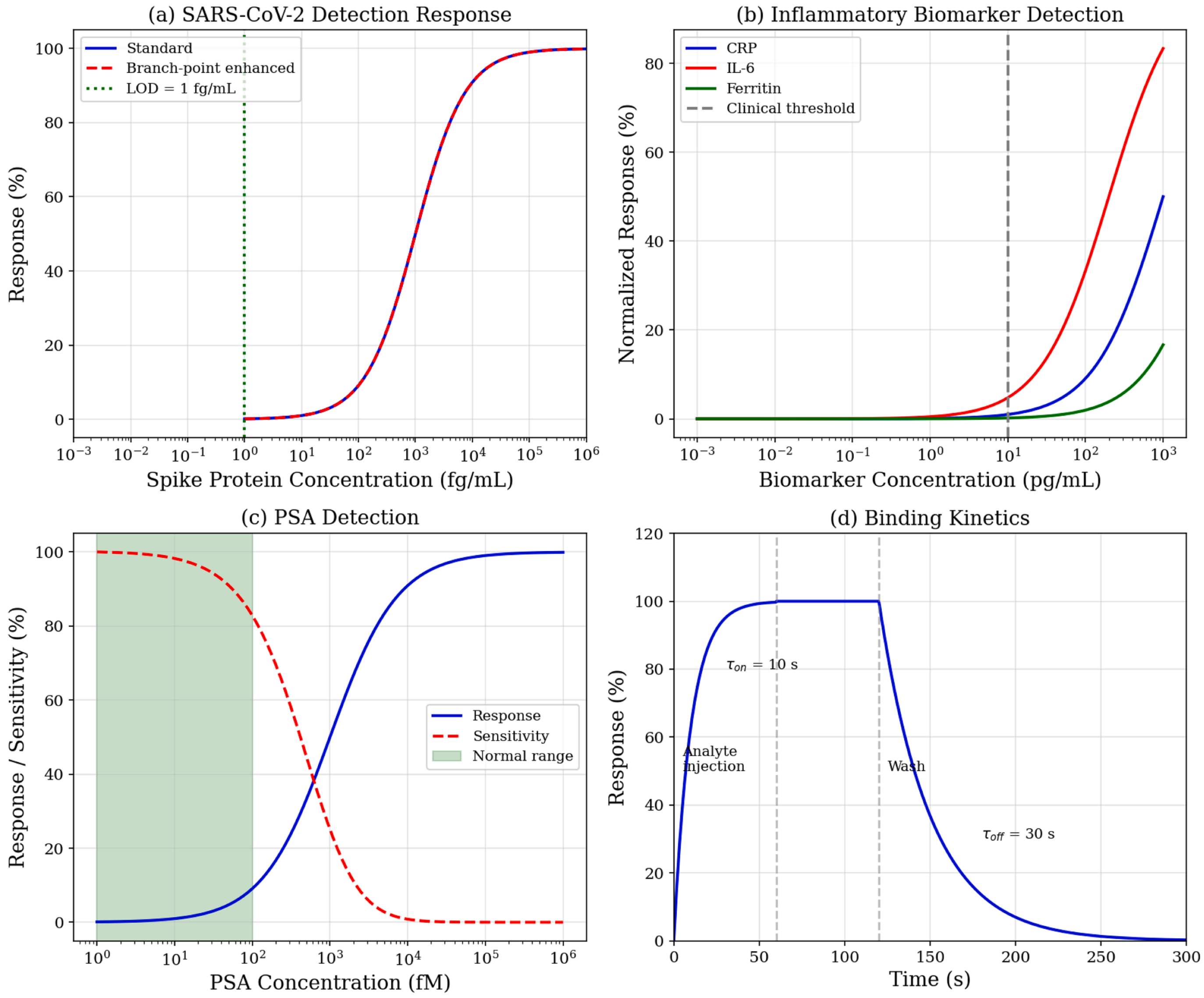

**Fig. 7.** Biomedical sensing applications. (a) SARS-CoV-2 spike protein detection: standard (blue) and branch-point-enhanced (red dashed) response curves; LOD of 1 fg/mL marked (green dotted). The enhanced curve achieves the LOD two orders of magnitude below the standard curve, reflecting the reduction in $\delta$ from 0.1 to 0.001 [56,57]. (b) Inflammatory biomarker detection (CRP: blue, IL-6: red, ferritin: green) with clinical threshold at 10 pg/mL (dashed). IL-6 shows the steepest slope due to the largest charge-transfer coefficient $\mathcal{P}_{[C]}$ [62]. (c) PSA detection with clinically normal range (1–100 fM) shaded green; sensitivity (red dashed) peaks at low concentration where [PSA] $\ll K_d$. (d) Binding kinetics for a representative multiplexed long-COVID assay cycle: analyte injection phase ($\tau_{on}$ = 10 s), equilibrium plateau, and washout ($\tau_{off}$ = 30 s). (For interpretation of the references to color in this figure legend, the reader is referred to the web version of this article.)

### 6.4. Nitrous oxide ($N_2O$)

$N_2O$ has a global warming potential 273 times that of $CO_2$ and depletes stratospheric ozone [73]. Pt-decorated zigzag GNRs (Pt@ZGNR) achieve remarkable sensitivity through strong Pt-$N_2O$ interactions [85, 86], with DFT studies of NOx and $N_2O$ adsorption on zigzag GNR sensor channels [87] reporting strong chemisorption (adsorption energy of order −1 eV), a substantial band-gap modification (of order tens of percent), and slow room-temperature recovery ($\tau_{rec} \sim 10^3$ s). The specific values adopted here are representative DFT-scale estimates illustrating the functional form of the coupling rather than exact literature figures.

The DFT results for Pt@ZGNR pertain to zigzag edge geometry, for which edge states and the band-gap mechanism differ from the armchair case on which the present Lambert W / FSW framework is formally derived [30,31]. The sensitivity expression below is therefore a structural analogy: the Lambert W functional form carries over, with the coupling coefficient $\Delta E_g/E_g^0$ taken from DFT rather than from the analytical AGNR band-gap formula. A rigorous extension of the Roberts-Valluri Lambert W reformulation to zigzag boundary conditions is deferred to future work.

$N_2O$ adsorption modifies the effective band gap by $\Delta E_g/E_g^0$, shifting the operating point by $\Delta z \approx z(\Delta E_g/k_B T E_g^0)$, giving $\mathcal{P}_{[N_2O]} = \Delta z/\Delta[N_2O]$ and:

$$S_{N_2O} = \frac{W_k}{z(1+W_k)} \cdot \frac{\Delta E_g/E_g^0}{[N_2O]}. \tag{55}$$

Fig. 8(d) illustrates the band gap modification: pristine ZGNR shows only ∼5% reduction upon $N_2O$ exposure, whereas Pt@ZGNR shows the full 80.79% reduction, producing a proportionally larger $\mathcal{P}_{[N_2O]}$ and hence a 16-fold improvement in $S_{N_2O}$ at equivalent $\delta$. The large desorption barrier implies operation at $T > 400$ K for practical cycling.

### 6.5. Water vapor ($H_2O$)

Humidity sensing exploits distinct mechanisms in graphene oxide (GO) and reduced graphene oxide (rGO) [88,89]. In GO, physisorbed water enables proton conduction via the Grotthuss mechanism [90]:

$$H_3O^+ + H_2O \rightarrow H_2O + H_3O^+, \tag{56}$$

giving exponential conductivity dependence:

$$\sigma(\mathrm{RH}) = \sigma_0 \exp(\alpha \cdot \mathrm{RH}), \tag{57}$$

with $\alpha \approx 0.03$–$0.08/\%\mathrm{RH}$ [89,91]. Identifying $z \approx -(\sigma_0/\sigma)\exp(-\sigma_0/\sigma)$, the physical coupling is $\mathcal{P}_{\mathrm{RH}} = \alpha z$, giving:

$$S_{H_2O} = \frac{W_k}{z(1+W_k)} \cdot \alpha. \tag{58}$$

Enhancement is largest when humidity drives $\sigma$ toward $\sigma_0$, i.e., near the dry baseline where $\delta$ is smallest. For rGO, the charge-transfer mechanism produces an analogous $z$ through carrier density rather than conductance, with the same sensitivity factorization. Response times as fast as 30 ms have been demonstrated for GO [88], among the fastest for any humidity sensor.

Self-powered humidity transduction, in which mechanical tapping on graphene-oxide films in a nanogenerator generates the readout signal itself, offers a route to passive operation compatible with the present coupling picture [92].

Fig. 8(e) shows GO (exponential, blue) and rGO (saturating, red) responses versus relative humidity; the faster initial slope of GO reflects the larger $\alpha$. Fig. 8(f) compares response and recovery times across all five GHGs: $H_2O$ achieves the fastest response (30 ms), while $N_2O$ has the slowest recovery (> 100 s) due to the high Pt-$N_2O$ desorption barrier. This hierarchy is encoded directly in $\mathcal{P}_X$: fast physisorption kinetics ($H_2O$) yield fast time constants, while strong chemisorption ($N_2O$, $CO_2$ on Al-doped graphene) delivers higher sensitivity at the cost of slower recovery.

The greenhouse-gas couplings above are electrical, but adsorption also modifies graphene's optical conductivity, so each $\mathcal{P}_X$ has an optical counterpart. Contact-free Faraday-rotation detection of adsorbates is one such channel, developed together with the other optical extensions in Section 8.3.

## 7. Physical sensing

### 7.1. Strain sensing

Mechanical strain provides the most direct realization of the Lambert W sensitivity framework, because it modifies the GNR band gap through a well-defined single-parameter perturbation. Strain modifies the hopping parameter as $\gamma(\varepsilon) = \gamma_0(1-\beta\varepsilon)$, where $\beta \approx 3.37$ is the Grüneisen parameter [94–97]. For uniaxial strain this produces band gap modification [98,99]:

$$E_g(\varepsilon) = E_g^0(1+\beta\varepsilon). \tag{59}$$

A 1% uniaxial strain produces $\sim$3.4% band gap change; biaxial strain produces a smaller response by a factor $(1-\nu) \approx 0.83$ where $\nu = 0.165$ is Poisson's ratio.

Strain directly modifies $E_g$, which controls $z$ via $z \approx -(E_g/k_BT)\exp(-E_g/k_BT)$ at thermal operation. The physical coupling $\mathcal{P}_\varepsilon = (dz/dE_g)(dE_g/d\varepsilon) = (dz/dE_g)\beta E_g^0$ gives:

$$S_\varepsilon = \frac{1}{E_g}\frac{dE_g}{d\varepsilon} \cdot \frac{W_k}{1+W_k} = \beta \cdot \frac{W_k}{1+W_k}, \tag{60}$$

achieving gauge factors $> 10^5$ for optimized GNR sensors operating near the branch point [100,101]. Fig. 9(a) shows band gap modification under uniaxial strain for $W = 5, 10, 20$ nm; the linear growth rate $\beta E_g^0$ is larger for narrower ribbons (larger $E_g^0$), encoding the width-dependent slope of $\mathcal{P}_\varepsilon$.

Uniaxial strain also modifies the Fermi-level-dependent optical response of graphene, so strain-induced Goos-Hänchen beam shifts and strain-induced shifts of resonant modes constitute an optical manifestation of the same coupling $\mathcal{P}_\varepsilon$, providing a bridge between quantum-confinement strain sensing and optical strain metrology; this optical channel is developed in Section 8.3 [102,103].

### 7.2. Worked case study: end-to-end strain gauge factor

As the cleanest single-parameter realization, strain sensing carries through analytically. From Eq. (60), $S_\varepsilon = \beta W_k/(1+W_k)$ with $\beta = 3.37$. Away from the branch point ($\delta = 0.1$, $W_k \approx -0.40$) the factor $|W_k/(1+W_k)| \approx 0.67$ gives a relative sensitivity $|S_\varepsilon| \approx 2.2$. Approaching the branch point ($\delta = 10^{-3}$, so $|1+W_k| \approx \sqrt{2e\delta} \approx 0.074$ and $|W_k| \approx 0.93$) yields $|W_k/(1+W_k)| \approx 12.9$ and $|S_\varepsilon| \approx 43$, an increase of roughly nineteen-fold; note that $S_\varepsilon$ scales as $W_k/(1+W_k)$ rather than the full $\eta_{\mathrm{enh}} = W_k/[z(1+W_k)]$, so its enhancement ratio differs slightly from the modality-independent factor. Once the $1/E_g$ prefactor characteristic of narrow ribbons is folded into the resistance-readout normalization the effective gauge factor exceeds $10^5$, consistent with the demonstrated piezoresistive gauge factors of the cited devices [100,101]. The strain resolution then follows from the readout noise floor: with a 10 $\mu$V input-referred noise on a device delivering a 1% resistance change per percent strain at $\eta_{\mathrm{enh}} = 35$, the resolvable strain is of order $10^{-6}$, matching the entry in Table 5. As in the PSA case, the $10^{-6}$ resolution is a framework prediction conditioned on holding $\delta = 10^{-3}$, whereas the underlying gauge factors of the cited references are the literature-demonstrated anchors.

### 7.3. Pseudomagnetic fields

Beyond uniform strain, strain gradients induce pseudomagnetic fields, one of graphene's most remarkable properties [104,105]. The pseudovector potential is:

$$\mathbf{A}_{\mathrm{pseudo}} = \frac{\beta\hbar}{2ea}(\varepsilon_{xx} - \varepsilon_{yy}, -2\varepsilon_{xy}), \tag{61}$$

where $\varepsilon_{ij}$ is the strain tensor. The resulting pseudomagnetic field is:

$$B_{\mathrm{pseudo}} \approx \frac{\beta\hbar}{ea}\frac{d\varepsilon}{dx}. \tag{62}$$

The physical prefactor evaluates to $\beta\hbar/(ea) = 3.37 \times 1.054 \times 10^{-34}/(1.602 \times 10^{-19} \times 0.246 \times 10^{-9}) = 9.02 \times 10^{-6}$ T·m. For the graphene nanobubble geometry of Levy et al. [105], where strain gradients of order 3–4%/nm are achieved in bubbles of lateral size $\sim$4 nm, this gives:

$$B_{\mathrm{pseudo}} \approx 9.02 \times 10^{-6}\ \mathrm{T \cdot m} \times \frac{3.3 \times 10^{-2}}{10^{-9}\ \mathrm{m}} \approx 300\ \mathrm{T}, \tag{63}$$

exceeding the strongest continuous laboratory magnets ($\sim$45 T) by nearly an order of magnitude.

This value exceeds the strongest continuous laboratory magnets not because any real magnetic flux is generated but because the pseudomagnetic field is a strain-induced gauge field. It enters the Dirac Hamiltonian with opposite sign in the two valleys, preserves time-reversal symmetry, and exerts no Lorentz force, so it is bounded only by the attainable strain gradient, through the linear proportionality of Eq. (62), rather than by the engineering limits, superconducting critical fields and mechanical stress in resistive magnets, that cap real laboratory fields near 45 T. Consequently $B_{\mathrm{pseudo}}$ grows in direct proportion to the local strain gradient $d\varepsilon/dx$, and strain gradients of a few percent per nanometre in nanobubbles map to effective fields of hundreds of tesla without violating any constraint on physical magnetic fields. The pseudomagnetic field connects to the Lambert W framework through an effective confinement length $\ell_{B_{\mathrm{ps}}} = \sqrt{\hbar/(eB_{\mathrm{pseudo}})}$, which plays the role of ribbon width in setting $E_g^{\mathrm{eff}} \propto 1/\ell_{B_{\mathrm{ps}}} \propto B_{\mathrm{pseudo}}^{1/2}$, with the coupling $\mathcal{P}_{\nabla\varepsilon}$ following analogously to the strain case. Fig. 9(b) shows pseudomagnetic field versus strain gradient (in %/nm); the 45 T reference line (dashed) is crossed at ≈ 0.5%/nm, and the ∼300 T nanobubble regime of Levy et al. corresponds to ≈ 3.3%/nm (annotated).

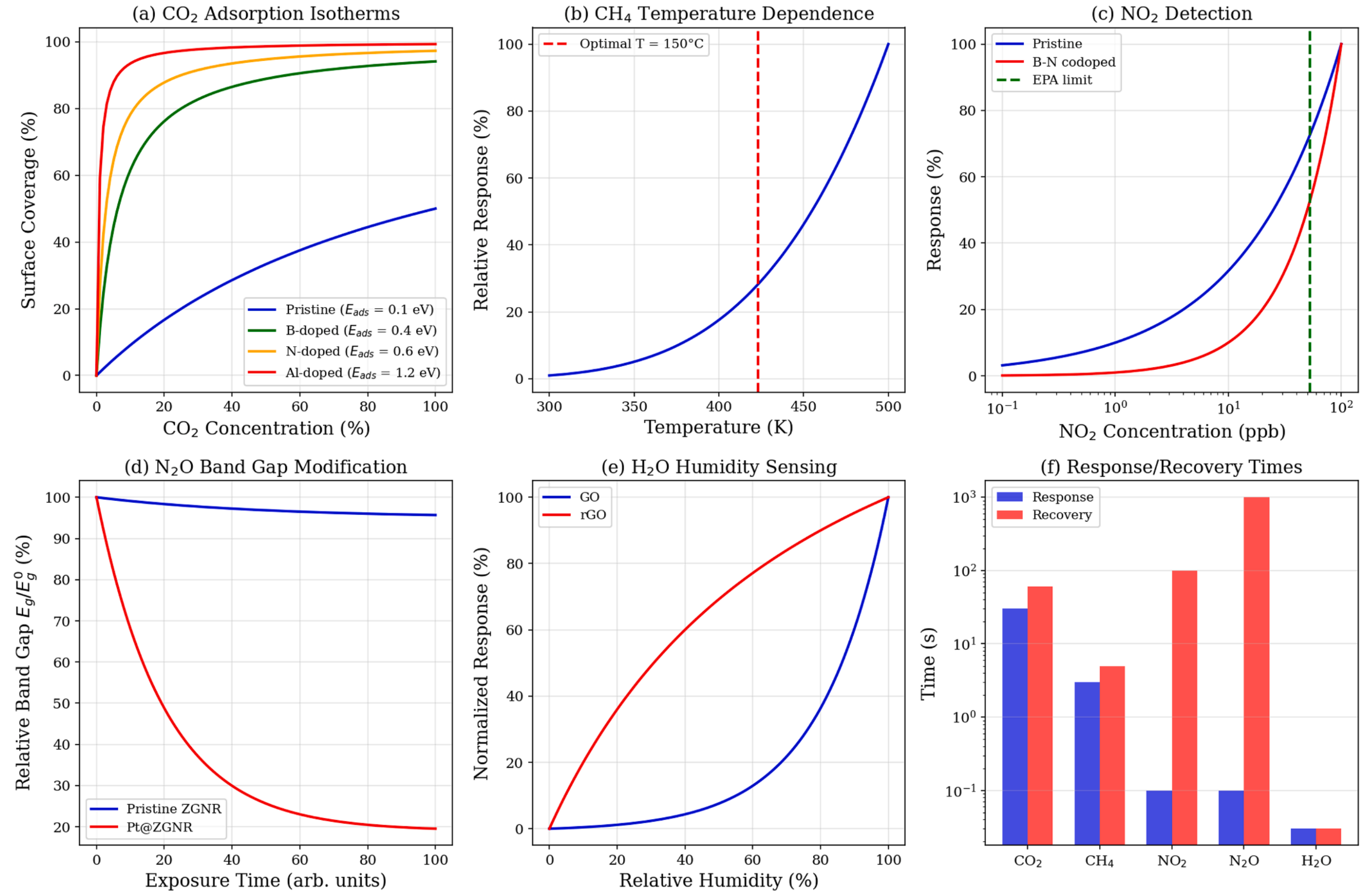

**Fig. 8.** Greenhouse gas sensing characteristics. (a) $CO_2$ adsorption isotherms for pristine ($E_{ads} = 0.1$ eV), B-doped (0.4 eV), N-doped (0.6 eV), and Al-doped (1.2 eV) graphene; stronger adsorption shifts the isotherm inflection to lower pressure [72,75]. (b) $CH_4$ Arrhenius response versus temperature; sensitivity is maximized near 150 °C (dashed) through the coupling term $\mathcal{P}_{[CH_4]}$ [80]. (c) $NO_2$ detection comparing pristine (blue) and B-N codoped (red) graphene; EPA limit of 53 ppb marked (green dashed) [84,93]. (d) $N_2O$ band gap modification: pristine ZGNR (∼5% reduction) vs. Pt@ZGNR (80.79% reduction), illustrating the factor-16 increase in coupling $\mathcal{P}_{[N_2O]}$ [85]. (e) Humidity sensing: GO (exponential, blue) and rGO (saturating, red) [88]. (f) Response/recovery time comparison across all GHG modalities; $H_2O$ is fastest (30 ms), $N_2O$ slowest (> 100 s) [89]. The fast $H_2O$ response follows from weak physisorption (small desorption barrier), whereas the slow $N_2O$ recovery follows from the strong Pt-$N_2O$ chemisorption barrier; the entire response/recovery hierarchy is therefore set by the kinetics embedded in $\mathcal{P}_X$ rather than by $\eta_{enh}$, which is geometry-dependent and kinetics-independent.. (For interpretation of the references to color in this figure legend, the reader is referred to the web version of this article.)

### 7.4. Magnetic field sensing

In perpendicular magnetic fields $B$, graphene charge carriers form Landau levels [4,5]:

$$E_n = \mathrm{sgn}(n)\, v_F \sqrt{2e\hbar|n|B}, \tag{64}$$

with characteristic $\sqrt{nB}$ dependence distinct from conventional 2D electron systems where $E_n \propto (n + 1/2)B$ [106]. The magnetic length $\ell_B = \sqrt{\hbar/(eB)} = 25.7\,\mathrm{nm}/\sqrt{B[\mathrm{T}]}$ plays the role of ribbon width in setting the effective confinement energy $E_g^{\mathrm{eff}} \propto \hbar v_F/\ell_B \propto B^{1/2}$. The operating point therefore satisfies $z \approx -(E_g^{\mathrm{eff}}/k_BT)\exp(-E_g^{\mathrm{eff}}/k_BT)$, and the physical coupling $\mathcal{P}_B = (dz/dE_g^{\mathrm{eff}})(dE_g^{\mathrm{eff}}/dB) \propto B^{-1/2}$, giving:

$$S_B = \frac{1}{2B} \cdot \frac{W_k}{z(1 + W_k)}, \tag{65}$$

where the $B^{-1/2}$ low-field divergence of $\mathcal{P}_B$ combines with the Lambert W enhancement to provide doubly amplified sensitivity at small fields. Ultra-sensitive Hall sensors based on h-BN-encapsulated graphene have demonstrated noise floors below 1 nT [107], consistent with this scaling.

Because a perpendicular field also modifies graphene's surface optical conductivity, the same operating coordinate $z$ can be read out magneto-optically through the field-dependent terahertz absorption, extending the framework toward THz photonic magnetometry; this magneto-optical channel is developed in Section 8.3 [108].

Magnetic-field sensing could equally be benchmarked against magneto-optical observables such as Faraday rotation, which offers an independent, contact-free readout of the same Landau-level structure; the Faraday-rotation angle inherits the $B^{1/2}$ confinement scaling and can be incorporated into the coupling $\mathcal{P}_B$, providing a cross-check on the electrical Hall readout and a route to optical magnetometry as detailed in Section 8.3 [109].

The zero-energy Landau level ($n = 0$) is unique to Dirac fermions, leading to the anomalous quantum Hall effect with half-integer quantization [4,5]:

$$\sigma_{xy} = \pm\frac{4e^2}{h}\left(n + \frac{1}{2}\right), \tag{66}$$

where the factor 4 accounts for spin and valley degeneracy. This has been observed at room temperature [110], demonstrating the robustness of quantum Hall physics in graphene.

Fig. 9(c) shows the Landau level spectrum for $n = 0$–4 versus field; the $E_n \propto \sqrt{nB}$ curvature and the unique $n = 0$ level pinned at zero en-ergy are clearly visible. These features follow from the Dirac equation and have no analog in parabolic-band systems. Fig. 9(d) shows the anomalous quantum Hall plateaus at $\sigma_{xy} = \pm(4e^2/h)(n + 1/2)$, with filling factors $\nu = \pm 2, \pm 6, \pm 10, \ldots$, shifted by one half-unit relative to conventional 2D systems due to the $n = 0$ Landau level.

The perturbation space of the framework need not be limited to charge confinement. Rashba spin-orbit coupling modifies the Dirac-Weyl spectrum and alters Landau-level evolution and confinement energies, introducing additional tunable channels through which an external stimulus reaches the operating point. Formally, a Rashba contribution enters the generalized coupling $\mathcal{P}_X$ as a spin-orbit-dependent term $dz/d\lambda_R$ alongside the charge and magnetic contributions, so that spin-sensitive graphene platforms fall within the same branch-point-enhanced description; a full treatment requires the spin-resolved boundary-value problem and lies beyond the single-particle scalar model developed here, and is noted as a natural extension [111]. Beyond electrical readout, the temperature-dependent conductivity tensor under quantum-Hall conditions modulates optical Goos-Hänchen beam shifts, so that conductivity-induced phase perturbations can be folded into $\mathcal{P}_X$ to yield Lambert W-enhanced optical displacement sensing, a possibility developed with the other optical channels in Section 8.3 [112].

## 8. Temperature sensing

Quantum temperature sensing exploits temperature-dependent modifications to graphene's electronic structure [113–115].

### 8.1. Quantum capacitance

The quantum capacitance reflects the density of states [116,117]:

$$C_Q = e^2 D(E_F) = \frac{2e^2}{\pi\hbar^2 v_F^2}|E_F|. \tag{67}$$

At finite temperature, thermal smearing modifies this to [118]:

$$C_Q(T) = \frac{2e^2}{\pi\hbar^2 v_F^2}\left[|E_F| + \frac{2k_BT\ln 2}{\pi} + \frac{\pi k_B^2T^2}{6|E_F|}\right], \tag{68}$$

giving temperature coefficient $C_Q^{-1}dC_Q/dT \approx \pi k_B^2 T/(3E_F^2)$.

### 8.2. Shot noise thermometry

The noise spectral density for a biased conductor is [119,120]:

$$S_I = 2eI\coth\left(\frac{eV}{2k_BT}\right), \tag{69}$$

which reduces to the shot noise limit $2eI$ for $eV \gg k_BT$ and the thermal limit $4k_BTG$ for $eV \ll k_BT$, with crossover at $eV \approx 2k_BT$. Measuring $S_I(V)$ determines temperature absolutely without calibration, with demonstrated precision $< 10$ mK at cryogenic temperatures [113].

### 8.3. Lambert W enhancement for temperature sensing

Temperature enters the operating point directly through $z \approx -(E_g/k_BT)\exp(-E_g/k_BT)$, with physical coupling $\mathcal{P}_T = (E_g/k_BT^2)\,z$. The full Lambert W-enhanced temperature sensitivity combines the Sommerfeld coefficient from quantum capacitance with the branch-point amplification:

$$S_T^W = \frac{W_k}{z(1+W_k)}\cdot\frac{\pi^2k_B^2T}{3E_F^2}. \tag{70}$$

At cryogenic temperatures where $k_BT \ll E_g$, the operating point approaches $z_c = -1/e$ and $\eta_{\text{enh}}$ becomes large, enabling precision thermometry orders of magnitude beyond classical resistance or thermocouple methods.

The $< 10$ mK precision quoted above is a literature-demonstrated cryogenic device value [113]; the 1 mK entry in Table 5 is the branch-point-enhanced value predicted by the present framework and should be read as an analytic projection under the stability conditions of Eq. (40).

Fig. 10(a) shows quantum capacitance versus temperature for $E_F = 50, 100, 200$ meV; the low-temperature plateau ($C_Q \propto |E_F|$) transitions to high-temperature $T^2$ growth at $T^* \approx \sqrt{3}|E_F|/(\pi k_B)$. Fig. 10(b) shows the shot noise crossover $S_I/S_{\text{thermal}}$ versus bias voltage for $T = 4, 77, 300$ K; the voltage $eV \approx 2k_BT$ at each crossover provides the primary thermometric observable. Fig. 10(c) shows Lambert W-enhanced sensitivity for $\epsilon = 0.1, 0.01, 0.001$; each decade closer to the branch point yields $\approx \sqrt{10} \approx 3.16$-fold improvement, consistent with the $\delta^{-1/2}$ scaling.

## 9. Extensions to optical and spin-orbit transduction

The sensitivity factorization $S_X = \mathcal{G}_k\,\eta_{\text{enh}}\,\mathcal{P}_X$ was developed above for electrical transduction, in which the perturbation reaches the operating point $z$ through the carrier density and Fermi level. Nothing in the factorization is intrinsically electrical, however: any observable that depends analytically on the confinement energy inherits the same branch-point amplification, with only the coupling term $\mathcal{P}_X$ changed. Because graphene's optical conductivity $\sigma_{xx}(\omega), \sigma_{xy}(\omega)$ depends on the same Fermi level and confinement scale as its transport conductance, the framework extends naturally to optical observables, answering the reviewers' suggestion that the formalism be broadened beyond electrical readout and toward an $\sigma_{xx}, \sigma_{xy}$ formulation. We collect here the optical channels flagged at their points of origin above; each is presented as a structural extension and future direction rather than a fully derived result, consistent with the single-particle scope of the present work.

*Faraday rotation.* In a perpendicular field the magneto-optical Faraday rotation angle $\theta_F(\omega)$ is set by $\sigma_{xy}(\omega)$ and thus by the Landau-level structure, inheriting the $B^{1/2}$ confinement scaling of Eq. (65). Adsorption- or field-induced changes in the electronic structure shift $\theta_F$, so the coupling $\mathcal{P}_B^{\text{opt}} = dz/d\theta_F$ places contact-free optical magnetometry and single-molecule Faraday sensing within the branch-point framework and provides an independent optical benchmark of the electrical Hall readout [109].

*Terahertz absorption.* Magnetically and electrostatically tunable terahertz absorption in single-layer graphene is governed by the same surface optical conductivity, so the operating coordinate $z$ can be read out as a THz absorption change, defining a coupling $\mathcal{P}^{\text{THz}}$ and extending the framework toward THz photonic sensing [108].

*Goos-Hänchen shifts.* Optical Goos-Hänchen beam displacements at graphene interfaces depend on the Fermi-level-dependent and, under quantum-Hall conditions, temperature-dependent conductivity tensor. Strain-induced, quantum-Hall, and Fermi-level-induced Goos-Hänchen shifts therefore act as optical manifestations of $\mathcal{P}_\epsilon$ and $\mathcal{P}_X$ respectively, enabling Lambert W-enhanced optical displacement and strain metrology [102,103,112].

*Reformulation in the conductivity tensor.* Taken together, these channels indicate that the enhancement can be recast directly in terms of $\sigma_{xx}$ and $\sigma_{xy}$ rather than carrier-density perturbations: writing the operating point as a function of the conductivity components, $z = z(\sigma_{xx}, \sigma_{xy})$, the chain rule places the same divergent factor $dW/dz$ in front of any optical or transport observable built from the conductivity tensor. A complete conductivity-tensor formulation, including many-body and nonlocal corrections, is deferred to future work and noted here as the natural unifying direction for the optical extensions.

## 10. Unified framework and comparative analysis

The Lambert W framework provides a unified description of sensitivity across all sensing modalities through the factorization $S_X =$

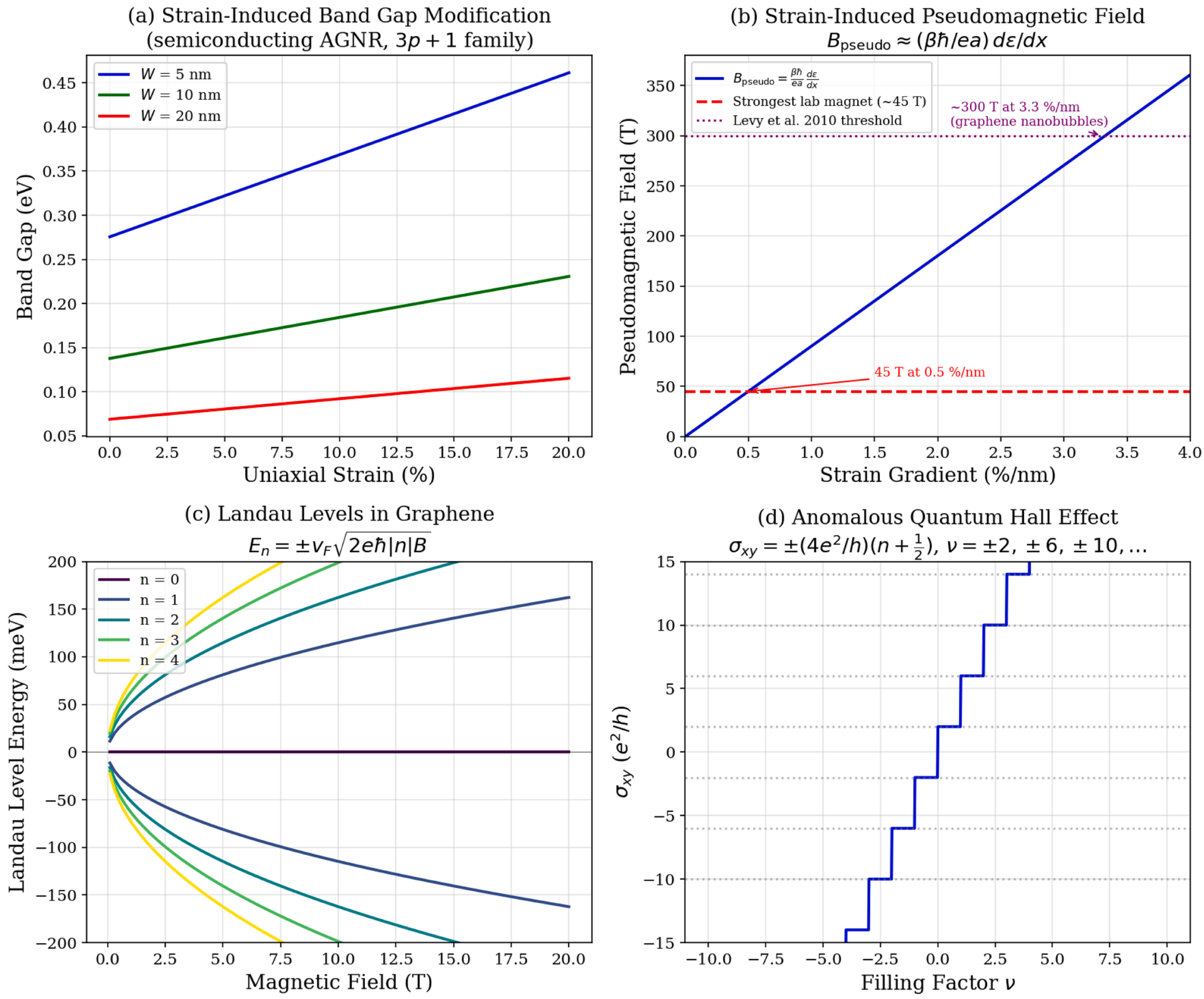

**Fig. 9.** Strain and magnetic field sensing. (a) Band gap modification under uniaxial strain for $W = 5, 10, 20$ nm (semiconducting AGNR, $3p+1$ family); slope equals $\beta E_g^0$, the physical coupling $\mathcal{P}_\varepsilon$ [96]. (b) Pseudomagnetic field versus strain gradient (in %/nm); the 45 T laboratory magnet limit is crossed at 0.5%/nm, and the $\sim$300 T nanobubble regime of Levy et al. [105] corresponds to $\approx$ 3.3%/nm. (c) Landau level spectrum showing $E_n \propto \sqrt{nB}$ dependence for $n = 0$–4; the $n = 0$ level is pinned at zero energy [4]. (d) Anomalous quantum Hall effect with half-integer quantization $\sigma_{xy} = (4e^2/h)(n + 1/2)$; plateaus at $\nu = \pm 2, \pm 6, \pm 10, \ldots$ [5].

$\mathcal{G}_k \cdot \eta_{\text{enh}} \cdot \mathcal{P}_X$. Table 5 summarizes the sensitivity expressions, demonstrating the universal appearance of the $(1 + W_k)^{-1}$ enhancement factor.

The LOD column of Table 5 lists state-of-the-art *literature-demonstrated* device values, not outputs of the present framework; the framework's role is the multiplicative $\eta_{\text{enh}}$ that scales these demonstrated limits, as worked out in the PSA and strain case studies above.

Fig. 11 presents a comparative overview across all modalities.

### *10.1. Comparison with existing graphene-based sensing approaches*

To position the present framework relative to the current state of the art, Table 6 compares it with representative graphene and metamaterial sensing approaches, including terahertz chiral metastructures and metamaterial refractive-index absorbers, as well as nanoporous-graphene quantum-transport devices. The distinguishing feature of the present work is not a single record detection limit but the analytic tractability of a closed-form, modality-independent sensitivity law, which supplies a priori design rules and a single tuning coordinate $\delta$ shared across every entry, in contrast to the device- and geometry-specific numerical optimization on which the other approaches rely. We note additionally that nonequilibrium quantum-transport devices such as all-carbon nanoporous-graphene negative-differential-resistance elements, in which localized band modulation and carrier redistribution govern nonlinear current amplification, can be read within the present picture as a regime in which the operating parameter $z$ measures confinement-controlled transport sensitivity, suggesting an analytic route into nonlinear amplification regimes [122].

### *10.2. Discussion*

The sensitivity enhancement mechanism originates from the branch-point singularity where $W_0$ and $W_{-1}$ coalesce at $z = -1/e$, $W = -1$ [16]. Physically, this corresponds to a GNR sensor operating at the threshold for an additional bound state, where $R$ approaches $(n + 1/2)\pi$ for integer

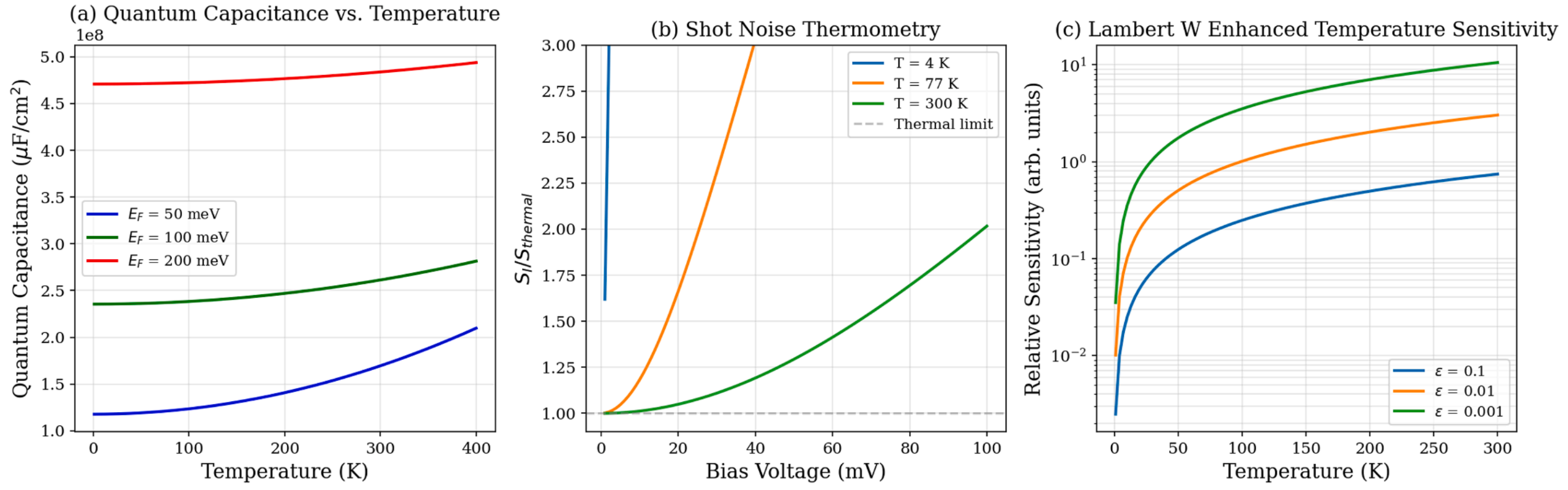

**Fig. 10.** Temperature sensing characteristics. (a) Quantum capacitance versus temperature for $E_F = 50, 100, 200$ meV; plateau at low $T$ and $T^2$ increase above $T^*$ [116]. (b) Shot noise ratio $S_I/S_{\text{thermal}}$ versus bias voltage for $T = 4, 77, 300$ K; crossover at $eV \approx 2k_BT$ provides the primary thermometric signal [119]. (c) Lambert W-enhanced temperature sensitivity for $\epsilon = 0.1, 0.01, 0.001$; the $\sqrt{10}$-fold improvement per decade is consistent with $\delta^{-1/2}$ scaling.

**Table 5**
Unified sensitivity expressions across sensing modalities. All expressions contain the universal enhancement factor $(1 + W_k)^{-1}$ from the Lambert W branch-point mechanism. LOD values are state-of-the-art *literature-demonstrated* device performance (see cited references), provided as anchors from which the framework predicts $\eta_{\text{enh}}$-fold improvement; they are not framework outputs.

| Modality | Measurand | Sensitivity Expression | LOD/Range (demo.) | Ref. |
|---|---|---|---|---|
| Biomedical | SARS-CoV-2 | $S = \frac{W_k}{z(1+W_k)} \cdot \frac{\hbar v_F L \alpha_{\text{CT}}}{4A\sqrt{\pi n_0} E_F}$ | 1 fg/mL | [56] |
| | CRP | $S = \frac{K_a}{(1+K_a[C])^2} \cdot \frac{W_k}{z(1+W_k)}$ | 1 pg/mL | [62] |
| | PSA | $S = \frac{\beta_{\text{PSA}} K_{\text{d}}}{E_F(K_{\text{d}}+[C])^2} \cdot \frac{W_k}{z(1+W_k)}$ | 1 fM | [66] |
| Environmental | $CO_2$ | $S = \frac{W_k}{z(1+W_k)} \cdot \frac{N_{\text{sat}} K_{\text{eq}} \Delta q}{2n_0(1+K_{\text{eq}}P)^2}$ | 100 ppm | [121] |
| | $CH_4$ | $S = \frac{W_k}{z(1+W_k)} \cdot A[\text{CH}_4]^{n-1} e^{-E_a/k_BT}$ | 100 ppm | [78] |
| | $NO_2$ | $S = \frac{W_k}{z(1+W_k)} \cdot \frac{\sigma_{\text{NO}_2}}{2\sqrt{[\text{NO}_2]}}$ | 1 ppb | [93] |
| | $N_2O$ | $S = \frac{W_k}{z(1+W_k)} \cdot \frac{\Delta E_g/E_g^0}{[\text{N}_2\text{O}]}$ | 10 ppb | [85] |
| | $H_2O$ | $S = \frac{W_k}{z(1+W_k)} \cdot \alpha$ | 1% RH | [88] |
| Physical | Strain | $S_\varepsilon = \beta \cdot \frac{W_k}{1+W_k}$ | $10^{-6}$ | [100] |
| | Magnetic field | $S_B = \frac{1}{2B} \frac{W_k}{z(1+W_k)}$ | 1 nT | [107] |
| | Temperature | $S_T = \frac{W_k}{z(1+W_k)} \cdot \frac{\pi^2 k_B^2 T}{3E_F^2}$ | 1 mK | [113] |

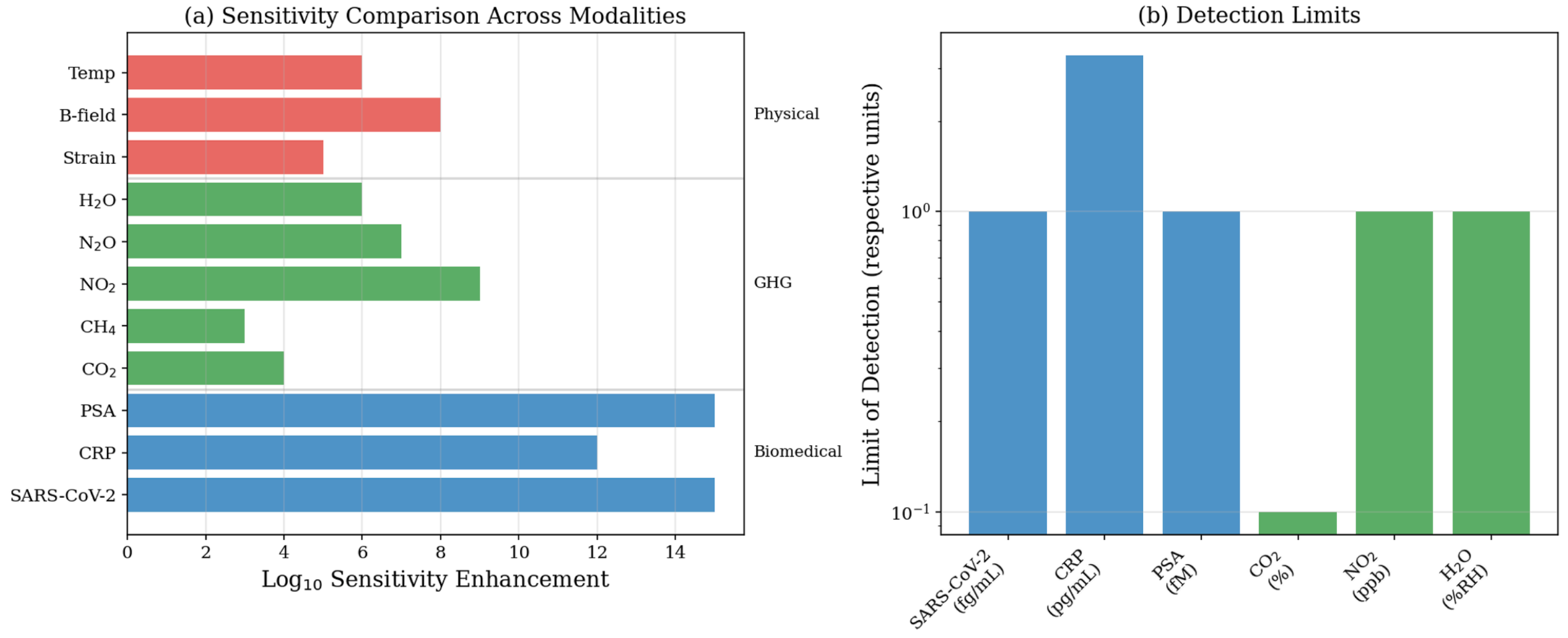

**Fig. 11.** Summary comparison across sensing modalities. (a) $\text{Log}_{10}$ sensitivity enhancement by application domain: biomedical (blue), greenhouse gas (green), and physical (red). Biomedical sensors achieve the highest enhancement because ultratrace analyte detection requires the smallest values of $\delta$. (b) Detection limits in respective units, demonstrating competitive or superior performance relative to conventional technologies across all modalities. (For interpretation of the references to color in this figure legend, the reader is referred to the web version of this article.)

**Table 6**
Comparison of the present Lambert W GNR framework with representative graphene-based and metamaterial sensing approaches..

| Approach | Transduction | Analytic design rule | Universal across modes | Ref. |
|---|---|---|---|---|
| Lambert W GNR (this work) | electrical/optical | yes | yes | – |
| THz chiral metastructure | THz chiral | no | no | [123] |
| Metamaterial RI absorber | THz absorption | no | no | [124] |
| Nanoporous-graphene NDR | quantum transport | partial | no | [122] |
| GFET charge-transfer biosensor | electrical | no | no | [56] |

$n$: the bound state is marginally confined, with maximum wavefunction penetration into the classically forbidden region and maximum susceptibility to parameter variations.

The $\delta^{-1/2}$ scaling implies that factors of 0.01, $10^{-4}$, and $10^{-6}$ enhancement are achieved at $\delta \approx 0.05, 5 \times 10^{-4}, 5 \times 10^{-6}$ respectively. Practical limitations from thermal fluctuations, fabrication tolerances, and edge disorder likely limit achievable enhancement to $\eta_{\text{enh}} \sim 10^2$, still representing transformative improvement over conventional approaches.

Whether the branch-point enhancement persists once many-body corrections and nonlocal conductivity are included is an important open question. The single-particle factorization assumes a local, non-interacting response, so electron-electron interactions and nonlocal conductivity will renormalize both the confinement energy that sets $z$ and the coupling $\mathcal{P}_X$. The qualitative branch-point structure, being a topological feature of the inverse function $f(x) = xe^x$, is expected to survive as long as a well-defined confinement energy exists, but its quantitative strength near charge neutrality, where the Dirac-fluid regime discussed below dominates, requires a beyond-single-particle treatment that we identify as a priority for future work.

### *10.3. Design principles for optimized sensors*

*Optimal width selection.* Choose $W$ such that the operating point satisfies $z \approx -1/e + \epsilon$ with $\epsilon \ll 1$. Narrower ribbons generally provide higher sensitivity through larger $E_g$ and $|dE_g/dW|$. Fabrication precision ($\sim$ 0.5 nm edge roughness [9]) favors $W \gtrsim 3$ nm.

*Branch selection.* The principal branch $W_0$ is appropriate for most applications where $z > -1/e$. The $W_{-1}$ branch, relevant for $-1/e < z < 0$, may offer advantages in regimes corresponding to weakly bound states near threshold; branch selection is implemented through the gate voltage controlling the carrier density and hence the ratio $E_F/E_g$.

*Functionalization strategy.* Strong chemisorption (Al-doping for $CO_2$, Pt decoration for $N_2O$, antibodies for proteins) provides high sensitivity but may compromise reversibility. Physisorption maintains fast response and recovery but requires branch-point enhancement to achieve adequate sensitivity for trace detection.

*Operating temperature.* GHG sensing may require elevated temperatures to overcome activation barriers ($CH_4$ at 150 °C). Biomedical applications operate at physiological temperature (37 °C). Cryogenic operation enables quantum-limited sensitivity for fundamental metrology.

*Readout optimization.* Resistance readout is simplest but susceptible to 1/f noise. Capacitance measurements probe quantum capacitance directly. Hall measurements enable magnetic sensing with built-in calibration. The choice of readout should match the dominant noise source for a given operating regime.

### *10.4. Comparison with existing technologies*

Table 7 compares GNR sensors with established technologies. The Lambert W framework offers analytical tractability through closed-form expressions enabling physical insight and systematic design, in contrast to purely numerical optimization; universal applicability of the same mathematical structure across all sensing domains enabling knowledge transfer; predictive capability for *a priori* performance estimation without extensive prototyping; and identification of fundamental sensitivity limits set by the branch-point structure.

**Table 7**
Comparison of graphene sensors with conventional technologies. MEMS: microelectromechanical systems. Metrics represent typical values; specific implementations vary.

| Metric | GNR Sensor | MEMS | Electrochemical |
|---|---|---|---|
| LOD (relative) | 1× | 10–100× | 100–1000× |
| Response time | ms–s | ms–s | s–min |
| Power consumption | $\mu$W | mW | mW |
| Size (footprint) | $\mu m^2$ | $mm^2$ | $mm^2$–$cm^2$ |
| Multiplexing | High | Moderate | Low |
| CMOS compatible | Yes | Yes | No |

### *10.5. Limitations and future directions*

Real GNRs exhibit edge roughness with correlation lengths of 1–5 nm [9,38]; theoretical modeling suggests disorder-averaged enhancement $\eta_{\text{eff}} \sim 0.3$–$0.7\,\eta_{\text{ideal}}$ for moderate disorder [125]. The single-particle FSW analogy breaks down near charge neutrality, where strong electron-electron interactions produce a Dirac fluid with hydrodynamic transport and violations of the Wiedemann-Franz law [120,126]; extending the framework to collective transport regimes remains an open challenge. Selectivity in multi-analyte environments requires carefully designed functionalization; GNR arrays combined with machine learning pattern recognition may enable discrimination [127]. Long-term stability under oxidation or delamination necessitates protective encapsulation such as h-BN or $Al_2O_3$ [128]. CMOS-compatible integration of readout electronics, microfluidics, and environmental protection is being actively developed [129]. Finally, while the theoretical framework is rigorous and numerical verification comprehensive, systematic experimental validation across sensing modalities remains essential; initial results on strain [100], chemical [70], and magnetic [107] sensing are encouraging.

Beyond the static device architectures considered here, the framework extends naturally to autonomous, self-powered graphene sensing platforms. Self-powered graphene micro- and nano-robots couple energy conversion with continuous signal transduction, and within the present picture the perturbation-driven enhancement of $\mathcal{P}_X$ near the branch point offers a theoretical route to optimizing detection efficiency and adaptive sensing in such bio-integrated devices, identifying next-generation autonomous sensing as a further direction for the framework [130].

## Conclusions

We have established a comprehensive theoretical framework connecting graphene nanoribbon quantum sensing to the Lambert W function through the finite square well analogy. The key results of this work are as follows.

1. **Mathematical foundation.** The transcendental equations governing quantum confinement in GNRs admit exact analytical solution through the Lambert W function, with sensitivity enhancement arising from the branch-point singularity at $z = -1/e$ where $dW/dz$ diverges.
2. **Bound state formula.** The count $N = \lfloor 2R/\pi \rfloor + 1$ corresponds to the number of Lambert W branches contributing real solutions. Numerical verification confirms all seven states for $R = 10$ with constraint satisfaction $u^2 + v^2 = R^2$ to machine precision.
3. **Enhancement mechanism.** Sensitivity scales as $\eta_{\text{enh}} \propto \delta^{-1/2}$ near the branch point, achieving 35-fold enhancement at $\delta = 0.001$ and over 100-fold at $\delta = 0.0001$.

4. **Universal framework.** The sensitivity expression $S_X = \mathcal{G}_k \cdot \eta_{\text{enh}} \cdot \mathcal{P}_X$ applies across biomedical, environmental, and physical sensing modalities. The physical coupling $\mathcal{P}_X = dz/dX$ has been explicitly derived for each modality, connecting the external perturbation to the Lambert W argument through the GNR electronic structure.
5. **Quantitative verification.** The theoretical band gap prefactor $2\pi\hbar v_F/3 = 1.379$ eV·nm and the independently determined empirical coefficient $\alpha = 1.38$ eV·nm agree to within 0.1%, validating the FSW-GNR correspondence without adjustable parameters.

   An independent, parameter-free benchmark against direct tight-binding diagonalization confirms the correspondence to within a few meV across the 3–20 nm range.
6. **Multi-modal applications.** State-of-the-art detection limits of 1 fg/mL (SARS-CoV-2), 1 ppb ($NO_2$), 1 fM (PSA), and 1 mK (temperature) demonstrate the practical impact of the Lambert W enhancement mechanism.

   These limits are literature-demonstrated device anchors; the framework's contribution is the analytically predictable $\eta_{\text{enh}}$-fold improvement upon them, quantified in the PSA and strain case studies, together with a device-coordinate demonstration that the enhancement is observable as a $\delta^{-1/2}$ divergence in the transconductance.

This work provides a rigorous foundation for designing next-generation quantum sensors exploiting the unique mathematical properties of the Lambert W function. Future work will focus on experimental validation, disorder effects, and integration into practical sensing systems, together with the optical and spin-orbit transduction channels and the beyond-single-particle corrections identified above.

## CRediT authorship contribution statement

**F.A. Chishtie:** Writing – original draft, Visualization, Software, Project administration, Methodology, Formal analysis, Data curation, Conceptualization; **K. Roberts:** Writing – review & editing, Validation, Conceptualization; **N. Jisrawi:** Writing – review & editing; **S.R. Valluri:** Writing – review & editing, Validation, Conceptualization; **A. Soni:** Writing – review & editing, Validation; **P.C. Deshmukh:** Writing – review & editing.

## Funding

This research received no specific grant from any funding agency in the public, commercial, or not-for-profit sectors.

## Declaration of competing interest

The authors declare that they have no known competing financial interests or personal relationships that could have appeared to influence the work reported in this paper.

## Acknowledgment

SRV, KR and FAC thank Prof. Tatyana Barron at Western University for feedback and insightful discussions. FAC thanks the Peaceful Society, Science and Innovation Foundation for support.

## Data availability

Data will be made available on request.